\documentclass{article} 
\usepackage{iclr2027_conference,times}

\usepackage{amsmath,amsfonts,bm}

\def\eqref#1{equation~\ref{#1}}

\def\1{\bm{1}}

\DeclareMathAlphabet{\mathsfit}{\encodingdefault}{\sfdefault}{m}{sl}
\SetMathAlphabet{\mathsfit}{bold}{\encodingdefault}{\sfdefault}{bx}{n}

\usepackage{hyperref}
\usepackage{url}
\usepackage{graphicx}
\usepackage{caption}
\usepackage{subcaption}
\usepackage{bm}
\usepackage{tabularx}
\usepackage{multirow}
\usepackage{makecell}
\usepackage{xr}
\usepackage{placeins}
\usepackage{enumitem}
\usepackage{array}
\usepackage{float}
\usepackage{dblfloatfix}
\usepackage{makecell}
\usepackage{amsmath}
\usepackage{algorithm}
\usepackage{algpseudocode}
\usepackage{wrapfig}
\usepackage{booktabs}
\newtheorem{theorem}{Theorem}[section]

\newtheorem{lemma}[theorem]{Lemma}
\usepackage{amssymb}
\usepackage{xcolor}
\usepackage{tikz}
\usetikzlibrary{arrows.meta,positioning,calc,fit}

\newcolumntype{C}[1]{>{\centering\arraybackslash}m{#1}}
\fontsize{9}{8}\selectfont

\title{Debiasing Message Passing \\to Mitigate Popularity Bias \\in GNN-based Collaborative Filtering}

\author{
Md Aminul Islam \quad
Ahmed Sayeed Faruk \quad
Sourav Medya \quad
Elena Zheleva \\
University of Illinois Chicago, Chicago, IL, USA \\
\texttt{\{mislam34,afaruk2,medya,ezheleva\}@uic.edu}
}

\iclrfinalcopy 
\begin{document}

\maketitle

\fancyhead{}
\fancyhead[L]{Preprint}

\begin{abstract}
    Collaborative filtering (CF) models based on graph neural networks (GNNs) achieve strong performance in recommender systems by propagating user-item signals over interaction graphs. However, they are susceptible to popularity bias, since skewed interactions and repeated message passing across high-order neighborhoods amplify the influence of popular items while suppressing long-tail ones. Existing debiasing approaches, including re-weighting objectives, regularization, causal methods, and post-processing, are less effective in GNN-based settings because they do not directly counteract bias propagated through the aggregation process, and recent in-aggregation weighting methods often rely on static heuristics or unstable embedding estimates. We propose \textit{Debiasing Popularity Amplification in Aggregation} (DPAA), a popularity debiasing framework for GNN-based CF that integrates adaptive, representation-aware interaction weighting and layer-wise weighting directly into message passing. DPAA assigns interaction-level weights from a representation-based popularity signal, stabilized by a smooth transition from pre-trained to evolving model embeddings during training. It further introduces a layer-wise weighting that amplifies higher-order neighborhoods, surfacing long-range interactions with diverse and underexposed items. Experiments on real-world and semi-synthetic datasets show that DPAA outperforms state-of-the-art popularity bias correction methods for GNN-based CF.
\end{abstract}

\section{Introduction} \label{sec:introduction}
Recommender systems have become a fundamental component of modern online platforms, helping users navigate large-scale information spaces and discover personalized content. These systems are deployed in diverse and complex digital ecosystems, including e-commerce and video-sharing platforms, social networks, and lifestyle applications~\citep{chen-inf23}. Collaborative filtering (CF) is a widely used recommendation paradigm that underpins many modern recommender systems. It infers user preferences directly from historical interaction data by leveraging patterns of similar user behavior~\citep{lubos-frontiers23}. In recent years, Graph Neural Networks (GNNs) have emerged as a powerful extension of traditional CF methods~\citep{he-sigir20, liu-www21, yu-kde23}. By representing user-item interactions as a bipartite graph, GNN-based CF models learn expressive representations through iterative neighborhood aggregation, capturing high-order collaborative signals, and achieving state-of-the-art recommendation performance~\citep{he-sigir20, yu-kde23}.

Recommender systems rely heavily on interaction data (e.g., clicks, views, watch events) to infer user preferences, but such data is often biased~\citep{ai-sigir18, joachims-wsdm17}, especially toward popular items~\citep{zhou-sigir23}. 
User interactions are not only driven by true preferences but are also influenced by exposure mechanisms, as users are more likely to interact with frequently displayed items~\citep{chen-inf23}. Hence, observed interactions typically follow a long-tail distribution~\citep{zhang-neurips22}, 
causing CF models trained on them to capture and reinforce popularity bias~\citep{canamares-sigir18, yao-neurips17}.
This problem is further amplified in GNN-based CF, where message passing repeatedly propagates signals from high-degree popular items across local and higher-order 
neighborhoods~\citep{zhou-sigir23, chen-front24}. Compared with sparsely connected niche items, popular items appear in many neighborhoods and exert disproportionate influence as their signals diffuse more broadly during message passing~\citep{chen-aaai20, chen-front24}. This repeated aggregation pulls user representations toward popularity-dominated regions, making popular items appear close to many users, increasing their predicted scores~\citep{zhou-sigir23}, and causing them to be recommended more frequently. Such dynamics create a ``rich-get-richer'' feedback loop~\citep{chaney-recsys18}, where popular items get more exposure while relevant long-tail items remain underexposed, degrading recommendation quality and user experience.

Existing popularity bias mitigation methods typically rely on inverse propensity weighting (IPW) to re-weight interactions~\citep{ai-sigir18, gruson-wsdm19, joachims-wsdm17}, regularization~\citep{boratto-ipm21, lin-wsdm25, liu-sigir23, zhang-neurips23}, causal modeling~\citep{bonner-recsys18, ning-www24, wang-kdd21, wei-sigkdd21, zhao-kde22}, or post-processing~\citep{abdollahpouri-arXiv19, chen-front24, jidoes-neurips25, zhu-kdd21} techniques. However, they do not address the repeated propagation and amplification of popularity bias during GNN aggregation~\citep{chen-front24, kim-cikm22, zhou-sigir23}.
To address this issue in GNN-based CF, recent studies~\citep{kim-cikm22, zhou-sigir23} incorporate weighting mechanisms directly into aggregation, assigning lower weights to interactions involving popular items, similar to propensity-based adjustments, to mitigate their disproportionate influence during message passing. For example, APDA~\citep{zhou-sigir23} estimates weights from the evolving user and item embeddings during training. However, early-epoch parameters can be noisy and unreliable due to rapid changes in weights and gradients~\citep{frankle-iclr20}. Since these noisy weights are applied during aggregation, they can distort representation learning by propagating erroneous signals into node representations in later epochs. NAVIP~\citep{kim-cikm22} weights aggregation using static interaction counts and ignores embedding-space popularity signals, limiting its ability to mitigate popularity amplification in learned representations. TSP~\citep{jidoes-neurips25} mitigates topology-induced bias after training to improve long-tail performance, but post-hoc correction cannot address how biased graph structure and repeated propagation of popularity signals shape representation learning. DAP~\citep{chen-front24}, a post-hoc GNN-based method, measures popularity using node degree, without considering interaction-level popularity effects encoded in the node embeddings.

We propose \textit{Debiasing Popularity Amplification in Aggregation} (DPAA), a popularity debiasing method for GNN-based CF that directly applies bias-correcting weights to user-item interactions during GNN aggregation. These weights downweight interactions during aggregation that are more likely to reflect popularity-induced interactions and upweight interactions involving niche items. We  introduce an embedding-aware interaction-strength metric that quantifies popularity-induced similarity in the embedding space at the interaction level and use it to assign weights during aggregation. To stabilize weight estimation in early training, the metric is initialized from a pre-trained model (e.g., the base recommendation model) and gradually transitions to the current model embeddings as they become reliable, yielding robust debiasing throughout training. To address popularity bias beyond the interaction level, we further introduce a layer-wise weighting mechanism in aggregation. 

Our main contributions are summarized as follows:
\begin{itemize}[leftmargin=10pt, nosep]
    \item \textbf{Direct bias correction. }We propose DPAA, a unified popularity debiasing method for GNN-based CF that directly mitigates popularity amplification during message passing. DPAA combines stabilized adaptive embedding-aware bias-correcting weights with layer-wise weights to address both interaction-level and layer-level sources of popularity  within a single method. DPAA controls whether interaction-level correction is applied at the initial or all layers, balancing under- and over-correction. Moreover, DPAA provides the flexibility to preserve node-specific information to mitigate over-smoothing. Therefore, DPAA differs from prior aggregation methods such as APDA and NAVIP in its weighting mechanism, stabilization of adaptive interaction weights, layer-wise weighting strategy, and control over possible over-correction and over-smoothing.
    \item \textbf{Theoretical characterization.} We show theoretically that our method DPAA reduces popularity dominance. More specifically, interaction-level weighting lowers the relative contribution of popular items depending on the severity of popularity bias, while layer-wise weighting increases the role of higher-order collaborative signals beyond shallow popularity-driven interactions.
    \item \textbf{Novel semi-synthetic data generation process.} Unlike prior studies that do not systematically evaluate debiasing methods under controlled popularity bias severity, we develop a semi-synthetic click data generation process that enables controlled variation of bias severity in the training data and allows us to evaluate how different methods perform for different levels of popularity bias.
    \item \textbf{Experiments.} Through experiments on real-world and semi-synthetic data, we show that DPAA outperforms existing methods from different methodological families of popularity bias correction across different bias levels and sparse scenarios, while remaining effective on the evaluated GNN-based CF backbones and improving performance for both popular and niche item groups.
\end{itemize}
\section{Related Work}
\label{app:related_work}
Existing approaches for mitigating popularity bias fall into several categories: re-weighting or IPW, causal inference, regularization-based methods, and post-processing methods. Re-weighting approaches, commonly known as IPW methods~\citep{ ai-sigir18, gruson-wsdm19, joachims-wsdm17, saito-recsys20, schnabel-icml16, yang-recsys18}, treat item popularity as a proxy for exposure probability and correct bias by weighting each instance inversely to that probability. To mitigate the high variance problem with such re-weighting, several studies~\citep{bottou-jmlr13, gruson-wsdm19} incorporate additional strategies, such as normalization techniques or smoothing regularization, to stabilize the training dynamics and produce more reliable estimates~\citep{bottou-jmlr13, gruson-wsdm19}. IPW has also been extensively adopted in learning-to-rank systems to address different biases, including position, selection, and popularity bias~\citep{joachims-wsdm17, oosterhuis-sigir20, wang-sigir16, ai-sigir18, luo-sigir24, luo-wsdm23, saito-recsys20, yuan-recsys20, wang-wsdm18}.

Regularization-based approaches~\citep{abdollahpouri-recsys17, boratto-ipm21, chen-sigir20, lin-wsdm25, liu-sigir23, zhu-wsdm21} address popularity bias by adding penalty terms that balance accuracy and coverage in recommendations. These methods differ in their penalty designs: Reg~\citep{zhu-wsdm21} separates item popularity from predicted user preferences; ESAM~\citep{chen-sigir20} employs attribute alignment, clustering, and self-training regularization to improve feature representations; ALS+Reg~\citep{abdollahpouri-recsys17} promotes fairness via intra-list diversity penalties; sam-reg~\citep{boratto-ipm21} reduces the biased correlation between relevance and item popularity, IPL applies a fairness criterion via a regularization term that ensures interactions are proportional to the number of users, and ReSN~\citep{lin-wsdm25} applies a spectral norm regularizer to penalize the popularity-induced principal singular value. Causal inference has been used to separate user interests from popularity effects~\citep{bonner-recsys18, he-icds22, ning-www24, wang-kdd21, wei-sigkdd21, zhang-sigir21, zhao-kde22, zheng-www21} by leveraging counterfactual reasoning to identify and reduce the impact of popularity on recommendations. CausE~\citep{bonner-recsys18} employs causal graphs for unbiased learning, MACR~\citep{wei-sigkdd21} removes direct popularity effects through multi-task counterfactual learning, MPCI~\citep{he-icds22} blocks paths from popularity to predictions, and PPAC~\citep{ning-www24} jointly models global and personal popularity within user neighborhoods.

Post-processing re-ranking methods~\citep{abdollahpouri-arXiv19, jidoes-neurips25, steck-recsys18, zhu-kdd21, zhu-wsdm21} typically operate on the generated recommendations without altering model parameters. Their objectives differ depending on the method: Calibration~\citep{steck-recsys18} preserves the historical preference distribution of users, RankALS~\citep{abdollahpouri-arXiv19} enhances recommendation diversity, and FPC~\citep{zhu-kdd21} addresses popularity bias in dynamic settings by adjusting predicted scores. Some studies have also investigated debiasing through out-of-distribution (OOD) generalization~\citep{wen-www22, zhao-www25-1, zhao-www25}, embedding landscape modeling~\citep{lee-cikm24}, and reinforcement learning approaches~\citep{lin-cikm22, shi-infs23}, with the goal of mitigating bias while maintaining a balance between recommendation accuracy and fairness. Moreover, CDN~\citep{zhang-kdd23} disentangles memorization from user behavior to alleviate long-tail bias in content-based recommendation systems.

GNNs~\citep{defferrard-neurips16, velivckovic-iclr18} have become a widely used method for capturing complex user-item interactions in recommender systems. Early models, such as NGCF~\citep{wang-sigir19}, employ iterative graph convolution to propagate collaborative signals, while more recent architectures, such as LightGCN~\citep{he-sigir20}, UltraGCN~\citep{mao-cikm21}, and SVD-GCN~\citep{peng-cikm20}, focus on enhancing computational efficiency and prediction accuracy. LightGCN~\citep{he-sigir20} removes feature transformations and non-linear activations to retain only linear neighborhood aggregation, UltraGCN~\citep{mao-cikm21} bypasses explicit message passing entirely using a constraint-based objective, and SVD-GCN~\citep{peng-cikm20} approximates graph convolution via truncated singular value decomposition for more efficient training. More recently, LightGCN++~\citep{lee-tors26} extends LightGCN with refined neighborhood aggregation and layer-wise representation combination to improve embedding quality and recommendation performance.

Incorporating contrastive and self-supervised learning~\citep{wang-cogmi21, wu-sigir21, yu-kde23, yu-sigir22, zhang-cikm21}, as in SimGCL~\citep{yu-sigir22} and XSimGCL~\citep{yu-kde23} has further strengthened representation quality through graph augmentations. DirectAU~\citep{wang-kdd22} emphasizes alignment and uniformity in representations, while SGL~\citep{wu-sigir21} and NCL~\citep{lin-www22} strengthen robustness by integrating contrastive learning with structural and semantic signals. GCF$_{logdet}$~\citep{zhang-neurips23} adds a regularization penalty during training to keep user and item embeddings diverse and spread out, preventing them from collapsing around popular items. Despite their effectiveness, these methods do not address the popularity amplification inherent in GNNs. APDA~\citep{zhou-sigir23} mitigates this by applying inverse popularity weights to interactions during aggregation, and NAVIP~\citep{kim-cikm22} leverages IPW based on static interaction counts during aggregation. DAP~\citep{chen-front24} is a post-hoc GNN-based debiasing method that estimates item popularity by clustering nodes with similar characteristics to remove the  popularity component from embeddings. TSP~\citep{jidoes-neurips25} mitigates topology-induced bias to improve long-tail item recommendations. Because TSP only applies corrections during inference, it cannot fully address bias that was baked into the embeddings during training. In contrast, our method directly intervenes during message passing, the source of bias amplification in GNNs, adjusting popularity-biased signals at the aggregation level. 
\section{Problem Formulation}
\textbf{Problem setup.} 
We consider a classical recommendation setting with users $\mathcal{U} = \{u_1, u_2, \dots, u_M\}$ and items $\mathcal{I} = \{i_1, i_2, \dots, i_N\}$. For each user-item pair $(u, i)$, let $y_{ui}$ be a binary indicator where $y_{ui}=1$ if user $u$ has interacted with item $i$ (e.g., through a click, rating, or watch event), and $y_{ui}=0$ otherwise. These interactions form a bipartite graph $\mathcal{G} = (\mathcal{V}, \mathcal{A})$, with node set  $\mathcal{V} = \mathcal{U} \cup \mathcal{I}$ ($|\mathcal{V}| = |\mathcal{U}| + |\mathcal{I}|$), and edges in $\mathcal{A}$ connecting users to the items they have interacted with. The goal is to recommend the top-$k$ most relevant items to each user that they have not interacted with.

\textbf{Background.} Graph convolutional networks (GCNs) have advanced representation learning in recommender systems~\citep{kipf-iclr17, mao-cikm21}. GNN-based CF approaches~\citep{he-sigir20, lin-www22, wang-sigir19, wang-sigir20, wu-sigir21, yu-kde23, yu-sigir22} are built on neighborhood aggregation, where node representations are iteratively refined by aggregating information from their neighbors. A single graph convolution operation can be expressed as:
\begin{equation} \label{eq:a_node_aggr}
    \begin{aligned}
        \mathbf{e}_u^{(l+1)} &= f_{\text{aggregate}}\Big(\{\mathbf{e}_i^{(l)} : i \in \mathcal{N}_u \cup \{u\}\}\Big), \quad
        \mathbf{e}_i^{(l+1)} &= f_{\text{aggregate}}\Big(\{\mathbf{e}_u^{(l)} : u \in \mathcal{N}_i \cup \{i\}\}\Big), 
    \end{aligned}
\end{equation}
where $\mathbf{e}_u^{(l)}$ and $\mathbf{e}_i^{(l)}$ represent the embeddings of user $u$ and item $i$ after $l$ propagation layers, while $\mathcal{N}_u = \{i \in \mathcal{I} : (u, i) \in \mathcal{A}\}$ and $\mathcal{N}_i = \{u \in \mathcal{U} : (u, i) \in \mathcal{A}\}$ denote the neighbor sets of user $u$ and item $i$. The aggregation function $f_{\text{aggregate}}(\cdot)$ combines information from neighboring nodes' previous-layer embeddings to produce updated representations. The initial embeddings $\mathbf{e}^{(0)}$ serve as trainable ID-based representations for each user and item. After $(l+1)$ rounds of message propagation, the user representation $\mathbf{e}_u^{(l+1)}$ or item representation  $\mathbf{e}_i^{(l+1)}$ captures information from its $(l+1)$-hop neighbors. By stacking $L$ propagation layers, a readout function $f_{\text{readout}}(\cdot)$ aggregates the representations from all layers to generate the final embedding:
\begin{equation} \label{eq:a_node_fianl_emb}
\begin{aligned}
    \mathbf{e}_u &= f_{\text{readout}}\Big([\mathbf{e}_u^{(0)}, \mathbf{e}_u^{(1)}, \dots, \mathbf{e}_u^{(L)}]\Big), \quad
    \mathbf{e}_i &= f_{\text{readout}}\Big([\mathbf{e}_i^{(0)}, \mathbf{e}_i^{(1)}, \dots, \mathbf{e}_i^{(L)}]\Big).
\end{aligned}
\end{equation}

Within GNN-based CF, LightGCN~\citep{he-sigir20} is a widely used backbone model. The model uses the GCN message passing paradigm by eliminating feature transformations and non-linear activations, retaining only the mechanism of linear neighborhood aggregation. It uses a weighted-sum aggregator as the aggregation function which is defined as: $\mathbf{e}_u^{(l+1)} = \sum_{i \in \mathcal{N}_u} \frac{1}{\sqrt{d_u d_i}} \mathbf{e}_i^{(l)}$,
$\mathbf{e}_i^{(l+1)} = \sum_{u \in \mathcal{N}_i} \frac{1}{\sqrt{d_i d_u}} \mathbf{e}_u^{(l)},$
where $d_u$ and $d_i$ denote the degrees of user $u$ and item $i$. To get the final representation, LightGCN employs a mean-pooling readout function that aggregates embeddings from all $(L+1)$ layers: $\mathbf{e}_u = \frac{1}{L+1} \sum_{l=0}^{L} \mathbf{e}_u^{(l)}$, 
$\mathbf{e}_i = \frac{1}{L+1} \sum_{l=0}^{L} \mathbf{e}_i^{(l)}.$
The predicted score between a user $u$ and an item $i$ is computed as the inner product of their final embeddings: $
\hat{y}_{ui} = \mathbf{e}_u^\top \mathbf{e}_i.$ LightGCN is optimized using the Bayesian Personalized Ranking (BPR) loss~\citep{rendle-arxiv12}. This is defined over triplets $(u,i,j)$, where $i$ is an interacted item for user $u$ and $j$ is not an interacted one, encouraging the model to predict $i$ higher than $j$:
\begin{equation} ~\label{eq:bpr_loss}
    \mathcal{L}_{BPR} = \sum_{u,i \in \mathcal{N}_u, j \notin \mathcal{N}_u} - \ln \sigma(\hat{y}_{ui} - \hat{y}_{uj}) + \rho \|\Theta\|_2^2.
\end{equation}
Here, $\rho$ is the regularization coefficient and $\Theta$ represents the current batch 0-th layer embeddings.

\textbf{Problem statement.} 
The goal of an unbiased GNN-based CF model is to learn node representations ($\mathbf{e}_u$ and $\mathbf{e}_i$) that are close to their optimal unbiased counterparts ($\mathbf{e}_u^{(*)}$ and $\mathbf{e}_i^{(*)}$). Unbiased representations are learned from interactions reflecting true preferences rather than bias factors. We can formalize the objective as minimizing the deviation between the model embeddings and their unbiased counterparts over all nodes $\mathcal{V}$:
$
\min_{\{\mathbf{e}_j\}_{j \in \mathcal{V}}}
\sum_{j \in \mathcal{V}}
\left\|
\mathbf{e}_j - \mathbf{e}_j^{(*)}
\right\|_2^2.
$
However, we cannot estimate the optimal unbiased embeddings because real-world data is biased. The challenge of debiasing is that we need to approximate unbiased representations from biased data. Our goal is to approximate unbiased embeddings indirectly by reducing the effect of bias during representation learning.
\section{Our Method: DPAA}
In this section, we introduce DPAA (\textit{Debiasing Popularity Amplification in Aggregation}), our proposed method for mitigating popularity bias in GNN-based CF. DPAA integrates adaptive bias-correcting weights directly into message passing. This is based on the insight that popularity bias is amplified during GNN aggregation. Therefore, effective mitigation requires intervening in the propagation stage. Formally, we modify the standard GNN message passing operator by introducing a weighted adjacency structure, where each user–item interaction is assigned a dynamic weight that depends on both interaction-level and layer-level factors. This results in a reweighted aggregation operator that explicitly controls the propagation of popularity-biased signals during training. 

\subsection{Estimating weights for counteracting popularity amplification}
\subsubsection{Interaction-level weight estimation}
In standard CF, the inner product between user and item embeddings measures user--item similarity and is used to rank items. The BPR objective encourages observed user--item pairs to have larger inner products than unobserved pairs. However, in popularity-biased data, observed positives are shaped by exposure. Popular items appear in many interactions, propagate their signals to a large number of users, and hence their embeddings become more similar to many user embeddings~\citep{zhou-sigir23}. Thus, a high inner product between a popular item's embedding and a user's embedding may reflect popularity-driven interactions~\citep{zhou-sigir23}. Based on this, we treat user--item similarity as a proxy for interaction strength, where highly similar user-item pairs are more likely to involve popular items. Note that symmetric normalization in graph aggregation (e.g., degree normalization in LightGCN) addresses graph structural imbalance but cannot alone prevent popular items from obtaining large inner products, as their many connections keep the sum of aggregated node degrees large, increasing the inner product~\citep{zhou-sigir23}. 

\textsc{Definition 1 (Current Model Inverse Interaction Weight).} \textit{Given the user and item embeddings $e_{u,t}^{(l)}$ and $e_{i,t}^{(l)}$ from the current model $c$ at layer $l$ and epoch $t$, we define the current model inverse interaction weight $r_{ui,c,t}^{(l)}$ as the complement of their $L_2$-normalized inner product:
\begin{equation}
    r_{ui,c,t}^{(l)} = r_{iu,c,t}^{(l)} = 1 - \frac{e_{u,t}^{(l)\top} e_{i,t}^{(l)}}{\|e_{u,t}^{(l)}\|_2 \, \|e_{i,t}^{(l)}\|_2}.
\end{equation}
}

\noindent Instead of the raw inner product, we use cosine similarity, which incorporates $L_2$ normalization to remove the effect of embedding magnitude.
$r_{ui,c,t}^{(l)}$ assigns weights inversely proportional to the interaction strength estimated from the current model embeddings, thereby giving smaller weights to interactions involving popular items and  larger weights to interactions involving long-tail items.

A naive approach that relies solely on current embeddings can be unstable, especially in early training, where embeddings change rapidly and are noisy~\citep{frankle-iclr20}. Applying weights estimated from such embeddings during aggregation may impact the learning process and embedding updates during backpropagation. To address this, we adopt an interaction-strength estimation strategy that uses embeddings from a pre-trained model. We use the embeddings of a pre-trained model, such as the base recommendation model that generated the current interaction data. This model provides stable embeddings for estimating weights at the early stages of training. These weights are pre-computed and remain fixed for each interaction at a specific layer $l$ throughout training.

\textsc{Definition 2 (Pre-trained Model Inverse Interaction Weight).} \textit{Given the user and item embeddings $e_{u,m}^{(l)}$ and $e_{i,m}^{(l)}$ from the pre-trained model $m$ at layer $l$, we define the pre-trained model inverse interaction weight $r_{ui, m}^{(l)}$ as the complement of their $L_2$-normalized inner product:
\begin{equation}
    r_{ui, m}^{(l)} = r_{iu, m}^{(l)} = 1 - \frac{e_{u, m}^{(l)\top} e_{i, m}^{(l)}}{\|e_{u, m}^{(l)}\|_2 \, \|e_{i, m}^{(l)}\|_2}.
\end{equation}}

To obtain an interaction-strength estimate that is reliable early in training and adaptive later, we combine the inverse interaction weights from the pre-trained and current models using a time-dependent convex combination. The pre-trained estimate provides stable signals in early training, while the current evolving model estimate captures evolving representations as learning progresses. This formulation ensures a smooth transition between the pre-trained and current model estimates. 

\textsc{Definition 3 (Inverse Interaction Weight).}
\textit{Given the current model inverse interaction weight $r_{ui, c, t}^{(l)}$ and the pre-trained model inverse interaction weight $r_{ui,m}^{(l)}$ at training epoch $t$, we define the inverse interaction weight (IIW) as:}
\begin{equation} \label{eq:final_iiw}
    r_{ui, t}^{(l)} = r_{iu, t}^{(l)} = \beta_t \, r_{ui,m}^{(l)} + (1 - \beta_t)\, r_{ui, c, t}^{(l)},
\end{equation}
\textit{where $\beta_t \in [0,1]$ is an epoch-dependent coefficient that controls the relative contribution of the pre-trained and current model estimates during training.}

The function is bounded in $[0,1]$ and adapts continuously based on embedding stability, avoiding abrupt transitions. In early epochs, $\beta_t$ is close to one, so IIW relies more on the pre-trained signal. As training progresses, $\beta_t$ decreases, allowing the current model to increasingly determine the weights.
To determine this transition, we measure the stability of current model embeddings using the average embedding change between consecutive epochs:
$\Delta_t
=
\frac{1}{|\mathcal{V}|}
\sum_{v \in \mathcal{V}}
\left\|
\mathbf{e}_{v,t} - \mathbf{e}_{v,t-1}
\right\|_2,$
where $\mathbf{e}_{v,t}$ and $\mathbf{e}_{v,t-1}$ denote the embeddings of node $v$ at epochs $t$ and $t-1$, respectively. A larger $\Delta_t$ indicates unstable embeddings, while a smaller $\Delta_t$ indicates more stable representations of the current model. We then define
$\beta_t
=
\frac{\Delta_t}{\Delta_t + C}$,
where $C \ge 0$ controls the sensitivity of the transition. With this formulation, large embedding changes yield $\beta_t \rightarrow 1$, making $r_{ui,t}^{(l)}$ rely mainly on the pre-trained estimate $r_{ui,m}^{(l)}$. As the embeddings stabilize and $\Delta_t$ decreases, $\beta_t \rightarrow 0$, so the current-model estimate $r_{ui,c,t}^{(l)}$ becomes dominant. This adaptive transition prevents noisy early-stage embedding fluctuations, as in~\citep{zhou-sigir23}, while preserving model-adaptive weighting in later training.

We provide empirical evidence that interactions with more popular items generally exhibit higher user--item similarity. By grouping items into different popularity quartiles, we observe that more popular item groups generally exhibit higher user–item cosine similarity in GNN-based CF models on real-world datasets (Figure 5 in the Appendix), potentially favoring popular items over tail items in recommendations from GNN-based CF models.
This supports our IIW mechanism, which uses the complement of cosine similarity to assign smaller weights during aggregation to interactions with higher user-item similarity. A detailed analysis is provided in Appendix~\ref{sec:popularity_similarity}.

\subsubsection{Layer-level weight estimation}
To address popularity bias beyond IIW, we introduce a layer-wise weighting mechanism that controls the contribution of different GNN layers during message passing. Lower layers mainly capture immediate user–item interactions, which are often dominated by frequently displayed items. In contrast, higher layers aggregate information from multi-hop neighbors, allowing user representations to incorporate signals from a more diverse set of items connected through shared collaborative contexts, including potentially relevant long-tail items. This enriches the representations beyond immediate popularity-driven interactions. Although deeper layers can carry popularity signals, their representations aggregate signals through multi-hop paths, making them less confined to a user’s immediate popularity-driven interactions. Therefore, we can assign larger weights to deeper layers, giving more importance to higher-order item signals and reducing shallow popularity dominance.

\textsc{Definition 4 (Layer-wise weighting).} 
\textit{We formalize the layer-wise weight (LW) as a function of the layer index. The weight assigned to layer $l$ is defined as:
\begin{equation}
    \lambda_l = (l + 1)^{\eta},
\end{equation}
where $\eta \geq 0$ is a hyperparameter controlling the strength of emphasis on different layers.}

In practice, $\lambda_l$ can be normalized across layers to maintain scale consistency and prevent domination by deep layers. When $\eta = 0$, all layers contribute equally, resulting in uniform aggregation. When $\eta > 0$, deeper layers receive larger weights, giving stronger emphasis to higher-order neighbors. The hyperparameter $\eta$ controls the contribution of deeper propagation, encouraging reliance on higher-order neighborhoods when beneficial while limiting over-reliance when they degrade performance.

The complete weighting mechanism unifies IIW and LW into a single formulation. The resulting weight \( w_{ui,t}^{(l)} \) for the interaction between user $u$ and item $i$ at layer \( l \) for epoch \( t \) can be written as:
\begin{equation} \label{eq:final_agg_weight}
w^{(l)}_{ui,t}
=
\lambda_l
\left[
(1-\gamma) r^{(l)}_{ui,t}
+
\gamma
\left(
\mathbb{I}(l=0) r^{(l)}_{ui,t}
+
\mathbb{I}(l>0)
\right)
\right].
\end{equation}
The parameter $\gamma \in \{0,1\}$ controls where IIW is applied, and $\mathbb{I}(\cdot)$ is an indicator function. When $\gamma=0$, IIW is applied at every propagation layer. When $\gamma=1$, IIW is applied only at the initial layer, while higher layers rely solely on LW. This design allows representations corrected by IIW at the initial layer to propagate to higher layers, while providing flexibility to apply IIW across all layers when deeper layers also require IIW-correction.
This helps balance potential under-correction when IIW is applied only at the initial layer and over-correction when it is applied across all layers.

\subsection{Debiasing popularity amplification in aggregation}
Embeddings at $(l+1)^{th}$ layer at training epoch $t$ can be written using the weight as follows:
\begin{equation} \label{eq:final_aggregation}
\begin{aligned}
\mathbf{e}_{u,t}^{(l+1)} = f_{\text{aggregate}}\Big(\{w_{ui,t}^{(l)}\mathbf{e}_{i,t}^{(l)} : i \in \mathcal{N}_u \cup \{u\}\}\Big),
\\
\mathbf{e}_{i,t}^{(l+1)} = f_{\text{aggregate}}\Big(\{w_{iu,t}^{(l)}\mathbf{e}_{u,t}^{(l)} : u \in \mathcal{N}_i \cup \{i\}\}\Big).
\end{aligned}
\end{equation}
\noindent Figure~\ref{fig:dpaa_overview} provides an overview of DPAA’s weighted message passing mechanism.
\definecolor{darktext}{HTML}{0F172A}
\definecolor{userblue}{HTML}{1746C7}
\definecolor{itemred}{HTML}{DC1717}
\definecolor{edgeorange}{HTML}{E89A24}

\definecolor{modelbg}{HTML}{EBF3FF}
\definecolor{modelborder}{HTML}{93B9EB}
\definecolor{iiwbg}{HTML}{EBF3FF}
\definecolor{iiwborder}{HTML}{93B9EB}
\definecolor{lwbg}{HTML}{EBF3FF}
\definecolor{lwborder}{HTML}{93B9EB}
\definecolor{weightbg}{HTML}{EBF3FF}
\definecolor{weightborder}{HTML}{93B9EB}

\tikzset{
    panel/.style={
        rounded corners=1.5mm,
        line width=0.7pt,
        inner sep=2mm,
        align=center
    },
    flow/.style={
        -{Latex[length=2.8mm]},
        draw=black!75,
        line width=1.2pt
    },
    user/.style={
        circle,
        fill=userblue,
        draw=white,
        line width=0.7pt,
        minimum size=0.48cm,
        inner sep=0pt
    },
    item/.style={
        circle,
        fill=itemred,
        draw=white,
        line width=0.7pt,
        minimum size=0.48cm,
        inner sep=0pt
    },
    interaction/.style={
        draw=edgeorange,
        line width=1.2pt,
        line cap=round
    },
    nodelabel/.style={
        font=\scriptsize,
        inner sep=1pt
    },
    weightlabel/.style={
        font=\fontsize{3}{3.4}\selectfont,
        text=black,
        fill=none,
        draw=none,
        inner sep=0pt
    }
}

\begin{figure*}[t]
\centering
\scalebox{1.0}[1.0]{%
\resizebox{\textwidth}{!}{%
\begin{tikzpicture}[
    x=1cm,
    y=1cm,
    every node/.style={text=darktext}
]


\node[font=\small\bfseries] at (4.9,5.75)
{Inverse interaction weight (IIW)};

\node[font=\small\bfseries] at (12.2,5.75)
{Layer-wise weight (LW)};


\node[
    panel,
    fill=modelbg,
    draw=modelborder,
    minimum width=4.7cm,
    minimum height=1.66cm,
    font=\scriptsize
] (prebox) at (2.45,4.60)
{
    {\small IIW from pre-trained model}\\[2mm]
    $\displaystyle
    r_{ui,m}^{(l)}
    =
    1-
    \frac{
        \mathbf e_{u,m}^{(l)\top}
        \mathbf e_{i,m}^{(l)}
    }{
        \|\mathbf e_{u,m}^{(l)}\|_2
        \|\mathbf e_{i,m}^{(l)}\|_2
    }
    $
};


\node[
    panel,
    fill=modelbg,
    draw=modelborder,
    minimum width=4.7cm,
    minimum height=1.66cm,
    font=\scriptsize
] (curbox) at (7.35,4.60)
{
    {\small IIW from current model}\\[2mm]
    $\displaystyle
    r_{ui,c,t}^{(l)}
    =
    1-
    \frac{
        \mathbf e_{u,t}^{(l)\top}
        \mathbf e_{i,t}^{(l)}
    }{
        \|\mathbf e_{u,t}^{(l)}\|_2
        \|\mathbf e_{i,t}^{(l)}\|_2
    }
    $
};


\node[
    panel,
    fill=lwbg,
    draw=lwborder,
    minimum width=4.2cm,
    minimum height=1.66cm,
    font=\small
] (lwcalc) at (12.2,4.60)
{
    $\displaystyle \lambda_l=(l+1)^\eta$
};


\node[
    panel,
    fill=iiwbg,
    draw=iiwborder,
    minimum width=9.6cm,
    minimum height=1.05cm,
    font=\small
] (iiw) at (4.9,2.45)
{
    Combined IIW\\[1.5mm]
    $\displaystyle
    r_{ui,t}^{(l)}
    =
    \beta_t r_{ui,m}^{(l)}
    +(1-\beta_t)r_{ui,c,t}^{(l)}
    $
};

\draw[flow]
    (prebox.south) -- (prebox.south |- iiw.north);

\draw[flow]
    (curbox.south) -- (curbox.south |- iiw.north);


\node[
    panel,
    fill=weightbg,
    draw=weightborder,
    minimum width=14.2cm,
    minimum height=1.25cm,
    font=\small
] (combined) at (7.25,0.35)
{
    Final message passing weight\\[2mm]
    $\displaystyle
    w_{ui,t}^{(l)}
    =
    \lambda_l
    \left[
        (1-\gamma)r_{ui,t}^{(l)}
        +
        \gamma
        \left(
            \mathbb{I}(l=0)r_{ui,t}^{(l)}
            +
            \mathbb{I}(l>0)
        \right)
    \right]
    $
};

\draw[flow]
    (iiw.south) -- (iiw.south |- combined.north);

\draw[flow]
    (lwcalc.south) -- (lwcalc.south |- combined.north);


\begin{scope}[shift={(19.1,0)}]

\node[font=\small\bfseries] at (0,5.75)
{Weighted message passing};

\begin{scope}[yshift=1.50cm,yscale=0.70]


\coordinate (U1) at (-2.00,1.00);
\coordinate (U2) at (-1.00,2.65);
\coordinate (U3) at ( 1.00,2.65);
\coordinate (U4) at ( 2.00,1.00);

\coordinate (I1) at ( 0.00,0.70);
\coordinate (I2) at (-3.00,4.30);
\coordinate (I3) at ( 2.10,4.70);
\coordinate (I4) at (-0.50,5.00);
\coordinate (I5) at ( 3.30,3.70);


\draw[interaction] (U1) -- (I1);
\draw[interaction] (U2) -- (I1);
\draw[interaction] (U3) -- (I1);
\draw[interaction] (U4) -- (I1);

\draw[interaction] (U1) -- (I2);
\draw[interaction] (U3) -- (I3);
\draw[interaction] (U2) -- (I4);
\draw[interaction] (U3) -- (I5);
\draw[interaction] (U3) -- (I4);
\draw[interaction] (U4) -- (I5);


\node[user] (u1) at (U1) {};
\node[user] (u2) at (U2) {};
\node[user] (u3) at (U3) {};
\node[user] (u4) at (U4) {};

\node[item] (i1) at (I1) {};
\node[item] (i2) at (I2) {};
\node[item] (i3) at (I3) {};
\node[item] (i4) at (I4) {};
\node[item] (i5) at (I5) {};


\node[weightlabel] at (-1.05,0.50)
    {$w_{u_1i_1,t}^{(l)}$};

\node[weightlabel] at (-1.00,1.45)
    {$w_{u_2i_1,t}^{(l)}$};

\node[weightlabel] at (1.05,1.45)
    {$w_{u_3i_1,t}^{(l)}$};

\node[weightlabel] at (1.05,0.40)
    {$w_{u_4i_1,t}^{(l)}$};

\node[weightlabel] at (-2.95,2.55)
    {$w_{u_1i_2,t}^{(l)}$};

\node[weightlabel] at (2.13,3.80)
    {$w_{u_3i_3,t}^{(l)}$};

\node[weightlabel] at (-1.28,3.85)
    {$w_{u_2i_4,t}^{(l)}$};

\node[weightlabel] at (0.69,4.05)
    {$w_{u_3i_4,t}^{(l)}$};

\node[weightlabel] at (2.30,2.80)
    {$w_{u_3i_5,t}^{(l)}$};

\node[weightlabel] at (3.09,2.18)
    {$w_{u_4i_5,t}^{(l)}$};


\node[nodelabel,text=userblue,anchor=east]
    at ([xshift=-0pt]u1.west) {$u_1$};

\node[nodelabel,text=userblue,anchor=east]
    at ([xshift=-0pt]u2.west) {$u_2$};

\node[nodelabel,text=userblue,anchor=north west]
    at ([xshift=0pt,yshift=-0pt]u3.south east) {$u_3$};

\node[nodelabel,text=userblue,anchor=west]
    at ([xshift=0pt]u4.east) {$u_4$};


\node[nodelabel,text=itemred,anchor=north]
    at ([yshift=-0pt]i1.south) {$i_1$};

\node[nodelabel,text=itemred,anchor=east]
    at ([xshift=-0pt]i2.west) {$i_2$};

\node[nodelabel,text=itemred,anchor=west]
    at ([xshift=0pt]i3.east) {$i_3$};

\node[nodelabel,text=itemred,anchor=west]
    at ([xshift=0pt]i4.east) {$i_4$};

\node[nodelabel,text=itemred,anchor=west]
    at ([xshift=0pt]i5.east) {$i_5$};

\end{scope}


\node[
    panel,
    fill=modelbg,
    draw=modelborder,
    minimum width=7.4cm,
    minimum height=1.25cm,
    font=\scriptsize,
    inner sep=1.5mm
] (updatedembeddings) at (0,0.47)
{
    {\small Updated embeddings at layer $l+1$}\\[2mm]
    $\displaystyle
    \begin{aligned}
    \mathbf{e}_{u,t}^{(l+1)}
    &= f_{\text{aggregate}}
    (\{w_{ui,t}^{(l)}\mathbf{e}_{i,t}^{(l)}
    :i\in\mathcal{N}_u\cup\{u\}\}), \\
    \mathbf{e}_{i,t}^{(l+1)}
    &= f_{\text{aggregate}}
    (\{w_{iu,t}^{(l)}\mathbf{e}_{u,t}^{(l)}
    :u\in\mathcal{N}_i\cup\{i\}\}).
    \end{aligned}
    $
};

\coordinate (graphwest) at (-3.80,2.10);

\end{scope}

\draw[flow]
    (combined.east)
    -- ++(0.45,0)
    |- (graphwest);

\end{tikzpicture}%
}
}

\caption{Illustration of the weighted message passing mechanism of DPAA. IIW estimates from the pre-trained and current models are combined for each user--item pair at epoch $t$, then combined with LW computed from the layer index to obtain final interaction weights. These weights are used in neighborhood aggregation to produce updated embeddings at layer $(l+1)$. Blue nodes denote users, red nodes denote items, and each interaction is annotated with its message passing weight.}
\label{fig:dpaa_overview}

\end{figure*}

When multiple GNN layers are stacked, over-smoothing~\citep{chen-aaai20} can make node representations  similar, reducing local structural distinctions among nodes. Since IIW is computed from user--item similarity, excessive smoothing can reduce variation among IIW values. In extreme case, when all  embeddings become nearly identical, IIW values also become nearly uniform, diminishing the bias-correcting capability of IIW. Moreover, assigning larger weights to deeper layers through LW can sometimes exacerbate over-smoothing. To mitigate these, following~\citep{chen-icml20}, we use an \textit{initial residual connection} $\delta$ that preserves node-specific information at each layer:
\begin{equation} \label{eq:residual_emb}
   \bm{\hat{e}}_{u, t}^{(l)} = \bm{e}_{u, t}^{(l)} + \delta \bm{e}_{u, t}^{(0)},\quad 
   \bm{\hat{e}}_{i, t}^{(l)} = \bm{e}_{i, t}^{(l)} + \delta \bm{e}_{i, t}^{(0)}.
\end{equation}

\subsection{Theoretical analysis} \label{subsec:theoretical_analysis}
We provide theoretical analyses explaining how the two main components of DPAA reduce popularity dominance. First, IIW reduces the relative influence of popular items, applying larger correction when popularity bias is more pronounced. Second, LW increases the role of higher-order signals by giving them larger contributions to embedding updates during backpropagation, allowing the model to incorporate more diverse and relevant items beyond popularity-driven interactions.

\textbf{Relative message passing contribution reduction via IIW.} 
We analyze the effect of applying IIW during message passing. Setting $\lambda_l = 1$ in \eqref{eq:final_aggregation}, the IIW-only aggregation from item $i$ to user $u$ at layer $(l+1)$ becomes
$\mathbf{e}^{(l+1)}_{u,t} = \sum_{i \in \mathcal{N}(u)} r^{(l)}_{ui,t}\,{\mathbf{e}}^{(l)}_{i,t}$.
Since $r^{(l)}_{ui,t}$ is the complement of the normalized user-item similarity, popular items, which become broadly similar with many users through repeated propagation, receive smaller weights and contribute less to aggregation. We define the message passing contribution of item $i$ as the total weighted magnitude of the messages it sends to its neighbors:
$\mathcal{I}^{(l)}_{\mathrm{IIW}_i}
=
\sum_{u \in \mathcal{N}(i)}
r^{(l)}_{ui,t}\,
\left\|{\mathbf{e}}^{(l)}_{i,t}\right\|_2
=
d_i\,
\bar{r}^{(l)}_{i,t}\,
\left\|{\mathbf{e}}^{(l)}_{i,t}\right\|_2$,
where $d_i = |\mathcal{N}(i)|$ is the item degree and 
$\bar{r}^{(l)}_{i,t} = \frac{1}{d_i} \sum_{u \in \mathcal{N}(i)} r^{(l)}_{ui,t}$ is its average IIW. The standard GNN contribution is obtained by setting $r^{(l)}_{ui,t}=1$, giving
$\mathcal{I}^{(l)}_i=d_i\|\mathbf{e}^{(l)}_{i,t}\|_2$.
Because popular items are repeatedly propagated through many users, they become more broadly similar with user embeddings than long-tail items~\citep{zhou-sigir23}. Thus, for any popular item $p$ and long-tail item $q$, we have $\bar{r}^{(l)}_{p,t}<\bar{r}^{(l)}_{q,t}$, with $\bar{r}^{(l)}_{q,t}>0$. 
\begin{lemma}[Popularity influence reduction.]
\label{lemma:IIW_reduction}
For any popular item $p$ and long-tail item $q$ with $d_p > d_q$ and 
$\bar{r}^{(l)}_{p,t}<\bar{r}^{(l)}_{q,t}$, IIW reduces the relative message-passing contribution of the popular item by the following closed-form positive reduction:
\begin{equation}
\label{eq:iiw-reduction}
\frac{\mathcal{I}^{(l)}_p}{\mathcal{I}^{(l)}_q}
-
\frac{\mathcal{I}^{(l)}_{\mathrm{IIW}_p}}{\mathcal{I}^{(l)}_{\mathrm{IIW}_q}}
=
\left(1 - \frac{\bar{r}^{(l)}_{p,t}}{\bar{r}^{(l)}_{q,t}}\right)
\cdot
\frac{\mathcal{I}^{(l)}_p}{\mathcal{I}^{(l)}_q}
> 0.
\end{equation}
\end{lemma}
We provide the proof in Appendix~\ref{app:proof-prop1}. 
Lemma~\ref{lemma:IIW_reduction} has two important implications:
\begin{itemize}[leftmargin=10pt, nosep]
    \item \textit{First,} the reduction depends on the gap between the average IIW values of popular and niche items. A larger gap, which can arise when popularity bias is severe, indicates a clearer embedding-space distinction between popular and long-tail items and leads to a larger reduction in popular-item dominance compared to standard GNN. Similarly, a smaller gap leads to a smaller reduction.
    \item \textit{Second,} the reduction is amplified by the structural degree imbalance $d_p/d_q$, since $\mathcal{I}^{(l)}_p$ and $\mathcal{I}^{(l)}_q$ depend on item degrees. When popular items have many interactions and long-tail items have very few, this imbalance becomes large, so IIW applies a larger correction compared to the standard GNN. Thus, IIW adaptively mitigates bias based on the severity of popularity skew in the data.
\end{itemize}

\textbf{Influence of LW on representation learning.}
We now analyze how the LW component of DPAA reshapes representation learning.
In GNN-based CF, embeddings are updated through the gradient of the objective function, so applying LW during message propagation in the forward pass affects how the embeddings are updated during backpropagation. We analyze how LW influences the gradients of the loss with respect to the embeddings. Consider a training triple $(u,i,j)$, where user $u$ has interacted with item $i$ but not with item $j$. Let $\phi_v(\mathbf{e}^{(0)})$ be a function of the initial embeddings, and let $\psi_\ell(\mathbf{h}_v^{(\ell)})$ be a function of the layer-$\ell$ 
neighborhood aggregation of node $v$.
\begin{lemma}[Layer-wise influence on representation learning.]
\label{lemma:lw}
Applying layer-wise weighting during message passing, the gradient 
of the objective function $\mathcal{L}_{uij}$ with respect to embedding 
$\mathbf{e}_v$ for any node $v \in \{u, i, j\}$ satisfies:
\begin{equation}
    \frac{\partial \mathcal{L}_{uij}}{\partial \mathbf{e}_v} 
    \propto 
    \phi_v(\mathbf{e}^{(0)}) 
    + 
    \sum_{\ell=1}^{L} \ell^\eta \cdot \psi_\ell(\mathbf{h}_v^{(\ell)}).
\end{equation}
\end{lemma}
We provide the proof in Appendix~\ref{app:lw-analysis}. Lemma~\ref{lemma:lw} implies the following:
\begin{itemize}[leftmargin=10pt, nosep]
    \item \textit{First,} for any two layers $\ell_a$ and $\ell_b$ where $\ell_a <  \ell_b$, the deeper layer receives a larger gradient coefficient $(\ell_a)^\eta < (\ell_b)^\eta$ when $\eta > 0$, whereas in a standard GNN without LW, all layers contribute equally (i.e., $\ell^\eta = 1$ for all $\ell$ when $\eta = 0$). LW therefore structurally amplifies gradient influence toward deeper layers, reducing reliance on shallow popularity-dominated updates and distributing representation learning across a broader set of items from higher-order neighborhoods.
    \item \textit{Second,} as $\eta$ increases, the ratio $(\ell_b)^\eta / 
    (\ell_a)^\eta > 1$ grows, progressively amplifying the contribution of 
    deeper layers to embedding updates. In scenarios where immediate interactions are heavily dominated by popular items, larger $\eta$ can be particularly beneficial, as this shifts gradient influence 
    away from popularity-skewed local interactions toward higher-order 
    collaborative signals.
\end{itemize}
\section{Experiments} \label{sec:experiments}
We evaluate our method DPAA by answering the following research questions:
\begin{itemize}[leftmargin=9pt, nosep]
    \item \textbf{RQ1:} How effective is DPAA relative to existing methods for mitigating popularity bias?
    \item \textbf{RQ2:} How robust is DPAA across different levels of popularity bias severity?
    \item \textbf{RQ3:} How does DPAA perform across popular and niche item groups, and is there a trade-off?
    \item \textbf{RQ4:} Does DPAA remain effective when applied to a different GNN-based CF backbone?
\end{itemize}

\subsection{Experimental setup} \label{subsec:exp_setup}
\textbf{Datasets and evaluation metrics.} We conduct experiments on three real-world datasets: Coat~\citep{schnabel-icml16}, KuaiRec~\citep{gao-cikm22}, and Yahoo! R3~\citep{marlin-recsys09}. These datasets cover different domains: coat shopping, short-video recommendation, and music recommendation. We choose these datasets because they are among the few public datasets with unbiased test sets for unbiased evaluations. Coat and Yahoo! R3 provide test sets collected with random exposure, while the KuaiRec test set is collected with a near-full exposure setting. These datasets vary in scale, sparsity, and distributions, reflecting diverse real-world scenarios. Figure~\ref{fig:dataset_distribution} in the Appendix shows that a small fraction of items accounts for $80\%$ of training interactions in these datasets, highlighting the popularity skew in real-world data.
We use Recall@$k$, NDCG@$k$, and HR@$k$ as evaluation metrics. We report results over five independent runs with different random seeds. Detailed dataset statistics and distributions are provided in Appendices~\ref{app:dataset_deatils} and~\ref{sec:popularity_distributions}.
Details on unbiased evaluation and ranking protocols are in Appendix~\ref{app:unbiased_eval}.

\textbf{Semi-synthetic data generation with controlled popularity bias.}
A popularity bias correction method should remain effective with varying bias severity, but real-world datasets do not allow direct control over the degree of popularity. To the best of our knowledge, existing studies do not evaluate debiasing methods across different bias levels.  To enable systematic evaluation, we develop a semi-synthetic click-data generation process using the unbiased KuaiRec test set, which is collected under full-exposure settings. We sample training interactions with a Zipfian distribution~\citep{newman-physics05} and vary the skewness parameter from $s=0$ to $s=9$, where larger $s$ indicates stronger popularity. We refer to $s$ as the bias severity parameter. Details are provided in Appendix~\ref{app:syn_data_gen_details}.

\textbf{Baselines.}
We compare DPAA with popularity bias correction methods from major debiasing families, including GNN aggregation re-weighting methods APDA~\citep{zhou-sigir23} and NAVIP~\citep{kim-cikm22}, post-hoc embedding correction method DAP~\citep{chen-front24}, causal-inference methods PPAC~\citep{ning-www24}, MACR~\citep{wei-sigkdd21} and CVIB~\citep{wang-neurips20}, IPW-based method IPS~\citep{gruson-wsdm19}, and regularization-based method SAM-REG~\citep{boratto-ipm21}. Since our study focuses on CF using only interaction data and no side information, we include baselines that are compatible with this setting. Unless stated otherwise, we use LightGCN as the backbone for our method and all baselines to ensure a fair comparison. We also use the more recent LightGCN++~\citep{lee-tors26} as the backbone for our method in RQ4 to validate its effectiveness with another GNN-based CF backbone. Detailed baseline descriptions and implementation details are provided in Appendix~\ref{app:baselines} and Appendix~\ref{app:impl_details}, respectively.

\subsection{Results} \label{subsec:results}

\begin{table*} [htpb]
   \fontsize{6.0}{6.0}\selectfont
   \setlength{\tabcolsep}{4.0pt}
   \newcommand{\stdsub}[1]{$_{\scalebox{0.7}{$\scriptscriptstyle\pm#1$}}$}
   \captionsetup{justification=raggedright, width=1.0\textwidth}
   \caption{Performance comparison of different methods. Bold and underline indicate the best and second-best results. Results are reported as mean $\bm{\pm}$ std\%. Improvement (\%) denotes the relative gain of DPAA over the best baseline.}
   \centering
    \begin{tabular}{cccccccccc}
    \toprule
     \multirow{2}{*}{Methods}
     & \multicolumn{3}{c}{Coat}
     & \multicolumn{3}{c}{KuaiRec}
     & \multicolumn{3}{c}{Yahoo! R3} \\
     \cmidrule(r){2-4} \cmidrule(r){5-7} \cmidrule(r){8-10}
        & Recall & NDCG & HR & Recall & NDCG & HR & Recall & NDCG & HR \\
    \midrule
    DPAA (Ours) & \textbf{0.2565}\stdsub{0.71} & \textbf{0.2065}\stdsub{0.87} & \textbf{0.6234}\stdsub{1.45} & \textbf{0.0349}\stdsub{0.49} & \textbf{0.0586}\stdsub{0.29} & \textbf{0.6201}\stdsub{1.02} & \textbf{0.1479}\stdsub{0.09} & \textbf{0.0659}\stdsub{0.06} & \textbf{0.2221}\stdsub{0.16} \\
    APDA & 0.1587\stdsub{0.25} & 0.0745\stdsub{0.51} & 0.4626\stdsub{0.69} & 0.0113\stdsub{0.62} & 0.0113\stdsub{0.41} & 0.1082\stdsub{1.22} & 0.1414\stdsub{0.09} & 0.0644\stdsub{0.60} & 0.2111\stdsub{0.42} \\
    NAVIP & 0.1255\stdsub{0.34} & 0.0635\stdsub{0.68} & 0.3808\stdsub{0.86} & 0.0016\stdsub{0.27} & 0.0090\stdsub{0.21} & 0.1128\stdsub{0.65} & \underline{0.1452}\stdsub{0.51} & \underline{0.0645}\stdsub{0.20} & \underline{0.2175}\stdsub{0.61} \\
    DAP & 0.1374\stdsub{0.77} & 0.0742\stdsub{1.02} & 0.4306\stdsub{1.68} & 0.0222\stdsub{0.72} & 0.0120\stdsub{0.46} & 0.1858\stdsub{1.12} & 0.1423\stdsub{0.21} & 0.0640\stdsub{0.18} & 0.2157\stdsub{0.24} \\
    PPAC & 0.1565\stdsub{0.49} & 0.0759\stdsub{0.61} & 0.4626\stdsub{1.27} & \underline{0.0313}\stdsub{0.23} & \underline{0.0474}\stdsub{0.24} & \underline{0.3340}\stdsub{0.76} & 0.1449\stdsub{0.11} & 0.0641\stdsub{0.09} & 0.2142\stdsub{0.21} \\
    MACR & 0.1002\stdsub{0.21} & 0.0451\stdsub{0.23} & 0.3345\stdsub{0.54} & 0.0039\stdsub{0.21} & 0.0041\stdsub{0.19} & 0.0816\stdsub{0.56} & 0.0095\stdsub{0.02} & 0.0033\stdsub{0.04} & 0.0190\stdsub{0.12} \\
    CVIB & \underline{0.2301}\stdsub{0.59} & 0.1880\stdsub{0.89} & \underline{0.5867}\stdsub{1.89} & 0.0116\stdsub{0.29} & 0.0134\stdsub{0.20} & 0.1106\stdsub{0.71} & 0.1435\stdsub{0.23} & 0.0643\stdsub{0.21} & 0.2127\stdsub{0.41} \\
    IPS & 0.2107\stdsub{0.19} & \underline{0.1905}\stdsub{0.23} & 0.5658\stdsub{0.48} & 0.0009\stdsub{0.23} & 0.0060\stdsub{0.20} & 0.0575\stdsub{0.49} & 0.1300\stdsub{0.31} & 0.0578\stdsub{0.19} & 0.1972\stdsub{0.35} \\
    SAM-REG & 0.1874\stdsub{0.81} & 0.1488\stdsub{0.78} & 0.5124\stdsub{0.91} & 0.0015\stdsub{0.29} & 0.0086\stdsub{0.31} & 0.0830\stdsub{0.71} & 0.1229\stdsub{0.53} & 0.0551\stdsub{0.30} & 0.1859\stdsub{0.89} \\
    LightGCN & 0.1301\stdsub{0.31} & 0.0715\stdsub{0.61} & 0.4093\stdsub{0.82} & 0.0039\stdsub{0.27} & 0.0029\stdsub{0.21} & 0.0487\stdsub{0.95} & 0.1411\stdsub{0.23} & 0.0628\stdsub{0.09} & 0.2134\stdsub{0.34} \\
    \cmidrule(r){1-10}
    Improvement (\%) & +11.47\% & +8.40\% & +6.26\% & +11.50\% & +23.63\% & +85.66\% & +1.86\% & +2.17\% & +2.11\% \\
    \bottomrule
    \end{tabular}
    \label{tab:main_results_unbiased}
\end{table*}
\textbf{Overall performance comparison (RQ1).}
We evaluate whether DPAA outperforms existing popularity bias correction methods and show the results in Table~\ref{tab:main_results_unbiased}. DPAA achieves the best performance on all datasets for Recall@20, NDCG@20, and HR@20. Compared with GNN aggregation-based baselines such as APDA and NAVIP, DPAA performs substantially better, suggesting the benefit of using stabilized IIW with LW. DPAA provides clear gains on Coat, which exhibits milder popularity skew. The largest improvements occur on KuaiRec, which exhibits severe popularity skew (Figure~\ref{fig:dataset_distribution}), suggesting that DPAA is especially effective when popularity strongly dominates the data. Some existing methods perform poorly on KuaiRec, likely due to its severe popularity skew, suggesting that they may be less suitable in such settings. On Yahoo! R3, DPAA outperforms all methods, although its gains are relatively small. This is likely due to the high sparsity of the dataset (Table~\ref{tab:dataset_stats}), which makes further improvements more challenging than on the other datasets for all methods. The low variability in the results suggests that DPAA's improvements over the baselines are consistent. These results demonstrate DPAA's effectiveness with varying popularity skew and sparsity. We analyze parameters in Appendix~\ref{app:add_results_rq1} and computational cost in Appendix~\ref{sec:computational_complexity}.

\textbf{Robustness under different popularity bias severity (RQ2).}
We vary the popularity bias severity $s$ from 0 to 9 using the semi-synthetic click generation process
on the KuaiRec dataset.
\begin{wrapfigure}{r}{0.47\columnwidth}
    \centering
    \includegraphics[width=0.47\columnwidth]{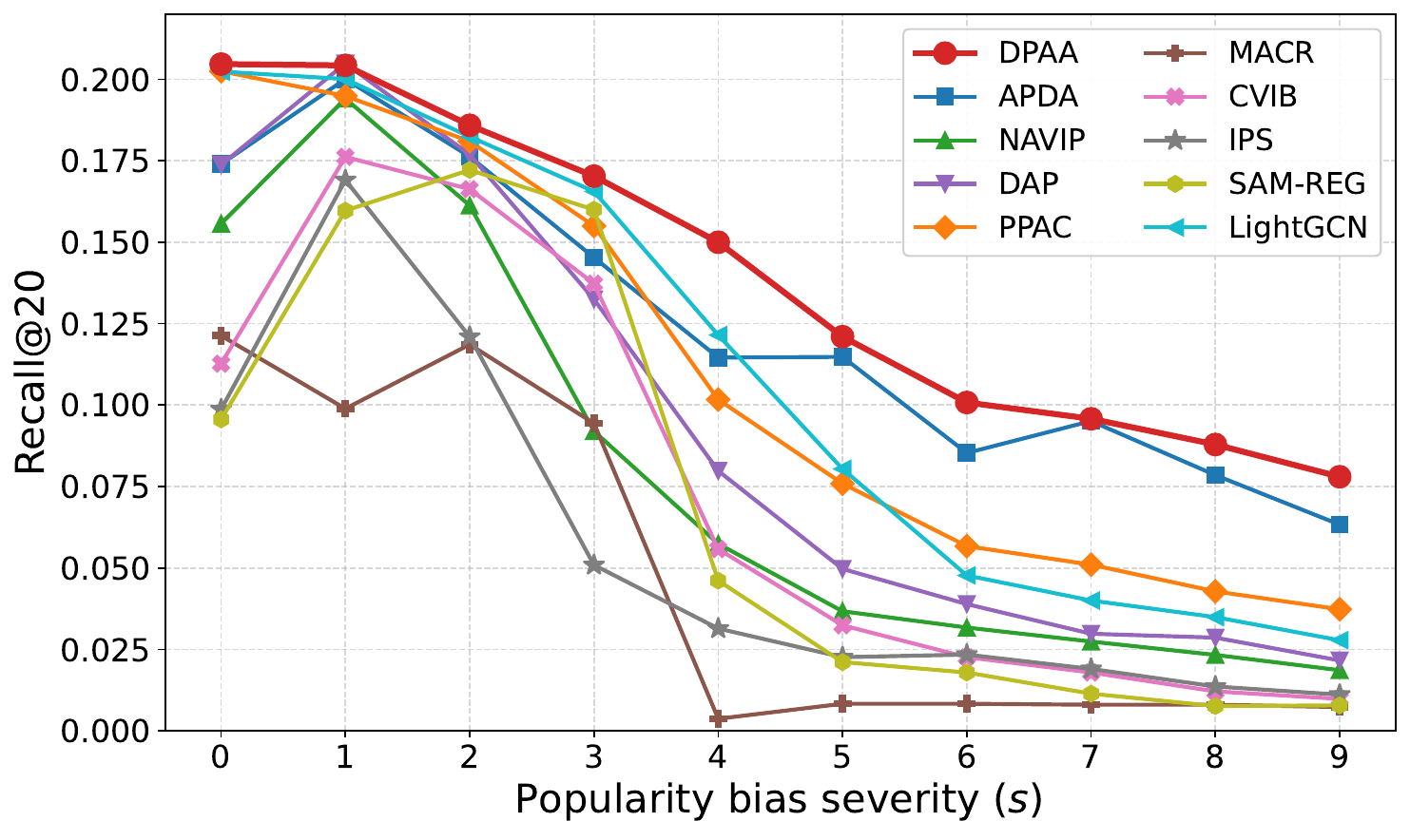}
    \captionsetup{justification=raggedright, margin=0cm}
    \captionsetup{width=0.47\textwidth}
    \caption{Performance comparison of different methods for Recall@20 on the KuaiRec dataset by varying levels of popularity bias severity ($s$).}
    \label{fig:severity}
\end{wrapfigure}
As shown in Figure~\ref{fig:severity}, DPAA outperforms all baselines except at $s=1$, achieving modest improvements at low bias severity ($s=0$--$3$) and larger gains as bias increases. When popularity bias is mild, LightGCN often outperforms most debiasing baselines, suggesting that some methods over-correct bias and hurt performance. DPAA remains competitive in these settings, indicating that it avoids over-correction. Although performance generally declines for all methods as bias increases, DPAA achieves larger gains over the baselines under severe bias, suggesting that it avoids under-correction. These results demonstrate DPAA's robustness across bias levels and are consistent with Lemma~\ref{lemma:IIW_reduction}, which shows that the reduction in popularity dominance depends on bias severity. We also provide the results in Table~\ref{tab:main_results_unbiased_synthetic} in the Appendix.

\textbf{Performance on popular and niche item groups (RQ3).}
RQ3 evaluates performance on popular and niche item groups, examining whether improving performance in one group affects the other. Following~\citep{abdollahpouri-arXiv19}, we sort items by interaction frequency and define the smallest set covering $80\%$ of interactions as popular items, with the remaining items treated as long-tail ones 
(Figure~\ref{fig:dataset_distribution}). As shown in Table~\ref{tab:popular_niche_split}, DPAA consistently outperforms the baselines across both item groups. The low KuaiRec niche scores reflect its highly imbalanced tail distribution, where many niche items have few interactions and may not closely align with users’ personalized preferences. As we perform unbiased evaluations, improvements in both groups show that DPAA does not simply shift recommendations from popular to niche items. It improves performance for both groups by reducing popularity amplification while not suppressing preference-relevant signals.

\begin{table*} [t]
   \fontsize{6.0}{6.0}\selectfont
   \setlength{\tabcolsep}{4.2pt} 
   \captionsetup{justification=raggedright, width=1.0\textwidth}
   \caption{Performance comparison on the popular and the niche item groups. Improvement (\%) denotes the relative gain of DPAA over the best baseline.}
   \centering
    \begin{tabular}{ccccccccccccc}
    \toprule
     \multirow{3}{*}{Methods} 
     & \multicolumn{4}{c}{Coat} 
     & \multicolumn{4}{c}{KuaiRec} 
     & \multicolumn{4}{c}{Yahoo! R3} \\
     \cmidrule(r){2-5} \cmidrule(r){6-9} \cmidrule(r){10-13}
     & \multicolumn{2}{c}{Popular} & \multicolumn{2}{c}{Niche}
     & \multicolumn{2}{c}{Popular} & \multicolumn{2}{c}{Niche}
     & \multicolumn{2}{c}{Popular} & \multicolumn{2}{c}{Niche} \\
     \cmidrule(r){2-3} \cmidrule(r){4-5} \cmidrule(r){6-7} \cmidrule(r){8-9} \cmidrule(r){10-11} \cmidrule(r){12-13}
        & Recall & NDCG & Recall & NDCG & Recall & NDCG & Recall & NDCG & Recall & NDCG & Recall & NDCG \\
    \midrule
    DPAA (Ours) & \textbf{0.3274} & \textbf{0.2288} & \textbf{0.1202} & \textbf{0.0917} & \textbf{0.0360} & \textbf{0.0601} & \textbf{0.0002} & \textbf{0.0003} & \textbf{0.2467} & \textbf{0.1064} & \textbf{0.0191} & \textbf{0.0065} \\
    APDA & 0.2097 & 0.0935 & 0.0057 & 0.0019 & 0.0128 & 0.0098 & 0.0000 & 0.0000 & 0.2329 & 0.1038 & 0.0155 & 0.0057 \\
    NAVIP & 0.1761 & 0.0818 & 0.0166 & 0.0053 & 0.0017 & 0.0064 & 0.0000 & 0.0000 & \underline{0.2383} & 0.1038 & \underline{0.0185} & \underline{0.0064} \\
    DAP & 0.1822 & 0.0870 & 0.0284 & 0.0090 & 0.0256 & 0.0134 & \underline{0.0001} & \underline{0.0001} & 0.1982 & 0.0864 & 0.0184 & 0.0060 \\
    PPAC & 0.1779 & 0.0918 & 0.0523 & 0.0223 & \underline{0.0331} & \underline{0.0477} & 0.0000 & 0.0000 & 0.2361 & 0.1016 & 0.0152 & 0.0059 \\
    MACR & 0.1141 & 0.0469 & 0.0555 & 0.0202 & 0.0045 & 0.0048 & 0.0000 & 0.0000 & 0.0003 & 0.0001 & 0.0183 & 0.0059 \\
    CVIB & \underline{0.2975} & \underline{0.2102} & 0.1093 & 0.0828 & 0.0274 & 0.0299 & 0.0000 & 0.0000 & 0.2372 & \underline{0.1043} & 0.0134 & 0.0057 \\
    IPS & 0.2643 & 0.2057 & \underline{0.1098} & \underline{0.0901} & 0.0012 & 0.0059 & 0.0000 & 0.0000 & 0.2120 & 0.0923 & 0.0180 & 0.0058 \\
    SAM-REG & 0.2242 & 0.1518 & 0.1093 & 0.0792 & 0.0018 & 0.0086 & \underline{0.0001} & \underline{0.0001} & 0.2096 & 0.0904 & 0.0066 & 0.0027 \\
    LightGCN & 0.1905 & 0.0939 & 0.0010 & 0.0005 & 0.0042 & 0.0027 & 0.0000 & 0.0000 & 0.2341 & 0.1017 & 0.0101 & 0.0044 \\
    \cmidrule(r){1-13}
    Improvement (\%) & +10.05\% & +8.85\% & +9.47\% & +1.78\% & +8.76\% & +26.00\% & +100.0\% & +200.0\% & +3.52\% & +2.01\% & +3.24\% & +1.56\% \\
    \bottomrule
    \end{tabular}
    \label{tab:popular_niche_split}
\end{table*}

\textbf{DPAA with a different GNN-based CF backbone (RQ4).}
We evaluate whether DPAA remains effective when used with another GNN-based CF backbone, such as LightGCN++~\citep{lee-tors26},
\begin{wrapfigure}{r}{0.47\columnwidth}
    \centering
    \includegraphics[width=0.47\columnwidth]{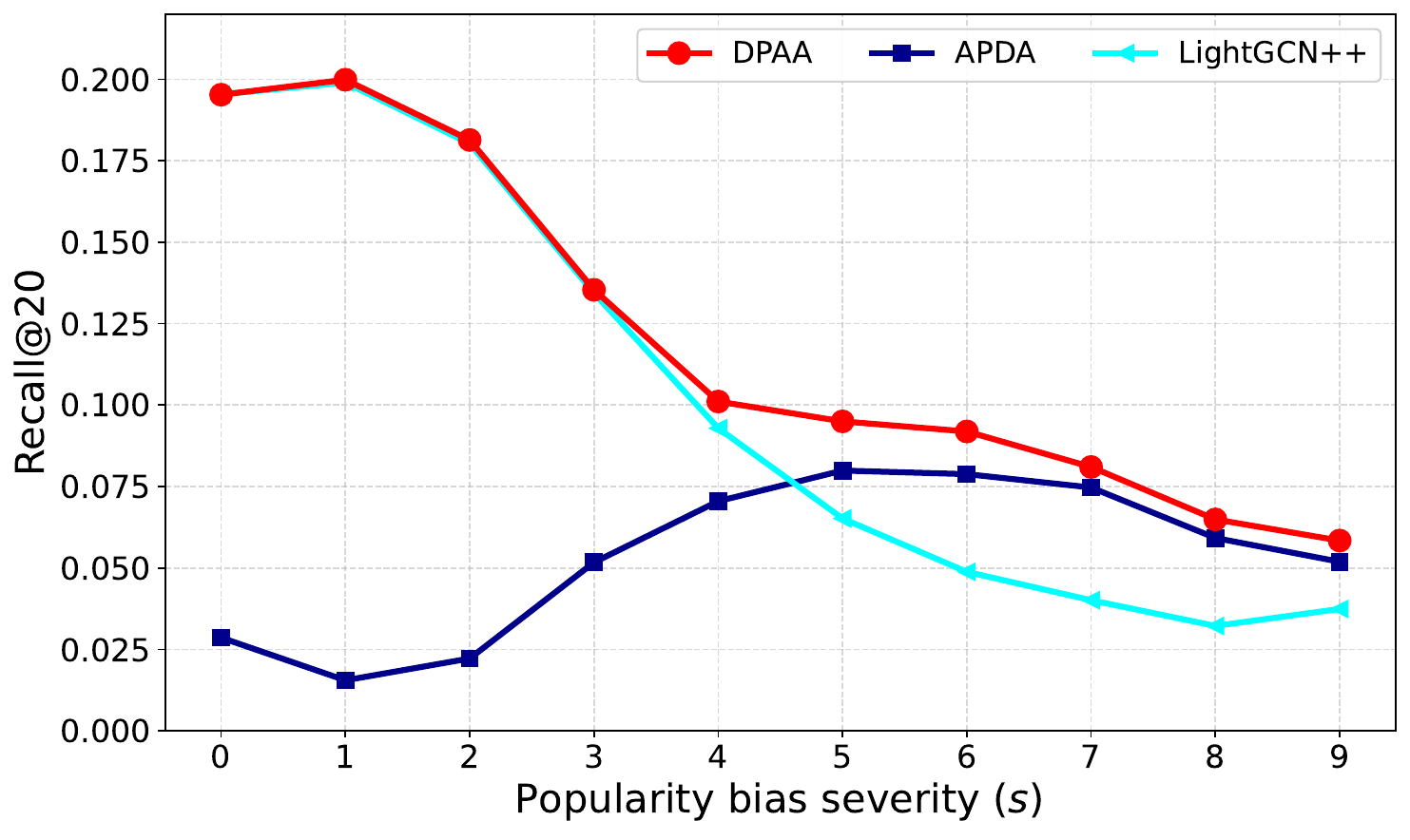}
    \captionsetup{justification=raggedright, margin=0cm}
    \captionsetup{width=0.47\textwidth}
    \caption{Performance using LightGCN++ as the backbone on the KuaiRec dataset.}
    \label{fig:severity_lightgcn++}
\end{wrapfigure}
under varying levels of popularity bias. We apply DPAA to the LightGCN++ backbone and compare it with the original LightGCN++, which does not correct for bias, and APDA, which is the best baseline on the KuaiRec semi-synthetic data. For a fair comparison, APDA is also applied to the LightGCN++ backbone. We show the Recall@20 results in Figure~\ref{fig:severity_lightgcn++} and report results for all three metrics in Table~\ref{tab:synthetic_bias_dpaa_lightgcn++} in the Appendix. DPAA outperforms across all bias severities from $s=1$ to $s=9$. DPAA performs comparably to LightGCN++ under mild bias, with substantially larger gains as severity increases. This also suggests that DPAA avoids over-correction and under-correction while effectively mitigating  popularity bias. These results suggest that, when applied to a different backbone, DPAA remains effective at mitigating popularity bias compared to other methods, especially with high bias severity.

\textbf{Ablation studies.}
We evaluate the contribution of each DPAA component. Table~\ref{tab:dpaa_variants} and Figure~\ref{fig:dpaa_variants} in the Appendix show Recall@20 on KuaiRec semi-synthetic click data with varying bias severity. Applying IIW to initial layers generally outperforms applying it to all layers, suggesting that aggressive correction may remove useful higher-order collaborative signals. For example, at bias severity levels 8 and 9, applying IIW only to the initial layer improves performance over applying it to all layers by 8.47\% and 4.81\%. Removing either the pre-trained or current model estimate degrades performance, while removing IIW or LW generally leads to further performance drops. For instance, at bias severity levels 8 and 9, DPAA improves Recall@20 by 22.01\% and 25.84\% over the variant without the current model for IIW estimation, by 18.67\% and 18.79\% over the variant without the pre-trained model, and by 40.28\% and 27.07\% over the variant without LW, respectively. These results show the importance of combining both models for IIW along with LW. Detailed component analyses are provided in Appendix~\ref{subsec:diff_components}. We also evaluate DPAA using pre-trained embeddings of varying quality (Table~\ref{tab:pretrain_sensitivity} in the Appendix). Better-quality pre-trained models generally improve performance, while lower-quality models cause only minor degradation, demonstrating DPAA's robustness to the choice of pre-trained model (Appendix~\ref{sec:pre-train_sensitivity}). We conduct hyperparameter ablations by varying $\delta$, $\eta$, and $C$. For $\delta$, smaller values perform best on Coat and KuaiRec, with $\delta=0.2$ yielding the best performance, while $\delta=0.6$ performs best on Yahoo! R3 (Figure~\ref{fig:residual_coefficient_variations_recall}). For $\eta$, the optimal value varies: $\eta=2$ for Coat, $\eta=3$ for KuaiRec, and $\eta=0$ for sparse Yahoo! R3 (Figure~\ref{fig:layer_weight_variations_recall}). For $C$, small nonzero values perform best on Coat and KuaiRec, while $C=0$ performs best on Yahoo! R3 (Figure~\ref{fig:C_variations_recall}). Detailed analyses are provided in Appendix~\ref{sec:ablation_parameter}.
\section{Conclusion} \label{sec:conclusion}
We propose DPAA, a popularity debiasing method for GNN-based CF that mitigates popularity amplification during message passing. DPAA combines adaptive interaction-level weighting with layer-wise weighting to reduce the influence of popular items during message passing and emphasize higher-order collaborative signals. Our theoretical analyses show that DPAA reduces the relative contribution of popular items while increasing the role of higher-order signals in representation learning. We also introduce a semi-synthetic click generation process to control popularity bias for systematic evaluation. Unbiased evaluations on real-world and semi-synthetic data show that DPAA outperforms existing methods, improves performance for both popular and niche items, and remains effective across the evaluated backbones and bias severities, with larger gains under severe bias.

\bibliography{main}
\bibliographystyle{iclr2027_conference}

\appendix
\clearpage
\section{Appendix}
\subsection{Theoretical Analysis} \label{app:theory_analysis}
\subsubsection{Proof of Lemma~\ref{lemma:IIW_reduction}}
\label{app:proof-prop1}

In the main body of the paper, we define the message-passing contributions as
$\mathcal{I}^{(l)}_i = d_i \cdot \|{\mathbf{e}}^{(l)}_{i,t}\|_2$ and
$\mathcal{I}^{(l)}_{\mathrm{IIW}_i} = d_i \cdot \bar{r}^{(l)}_{i,t} \cdot \|{\mathbf{e}}^{(l)}_{i,t}\|_2$,
where $\bar{r}^{(l)}_{i,t} = \frac{1}{d_i} \sum_{u \in \mathcal{N}(i)} r^{(l)}_{ui,t}$ is the average IIW of item $i$.

In standard GNN aggregation, the ratio of the message-passing contributions of a popular item $p$ and a long-tail item $q$ is:
\begin{equation*}
\frac{\mathcal{I}^{(l)}_p}{\mathcal{I}^{(l)}_q}
=
\frac{d_p \, \|{\mathbf{e}}^{(l)}_{p,t}\|_2}{d_q \, \|{\mathbf{e}}^{(l)}_{q,t}\|_2},
\end{equation*}
which is strictly positive since the degrees and embedding norms are positive. Under IIW, this ratio becomes:
\begin{equation*}
\frac{\mathcal{I}^{(l)}_{\mathrm{IIW}_p}}{\mathcal{I}^{(l)}_{\mathrm{IIW}_q}}
=
\frac{d_p \, \bar{r}^{(l)}_{p,t} \, \|{\mathbf{e}}^{(l)}_{p,t}\|_2}{d_q \, \bar{r}^{(l)}_{q,t} \, \|{\mathbf{e}}^{(l)}_{q,t}\|_2}
=
\frac{\bar{r}^{(l)}_{p,t}}{\bar{r}^{(l)}_{q,t}}
\cdot
\frac{\mathcal{I}^{(l)}_p}{\mathcal{I}^{(l)}_q}.
\end{equation*}

Subtracting the IIW-weighted ratio from the standard ratio gives:
\begin{align*}
\frac{\mathcal{I}^{(l)}_p}{\mathcal{I}^{(l)}_q}
-
\frac{\mathcal{I}^{(l)}_{\mathrm{IIW}_p}}{\mathcal{I}^{(l)}_{\mathrm{IIW}_q}}
&=
\frac{\mathcal{I}^{(l)}_p}{\mathcal{I}^{(l)}_q}
-
\frac{\bar{r}^{(l)}_{p,t}}{\bar{r}^{(l)}_{q,t}}
\cdot
\frac{\mathcal{I}^{(l)}_p}{\mathcal{I}^{(l)}_q} \\
&=
\left(1 - \frac{\bar{r}^{(l)}_{p,t}}{\bar{r}^{(l)}_{q,t}}\right)
\cdot
\frac{\mathcal{I}^{(l)}_p}{\mathcal{I}^{(l)}_q}.
\end{align*}

By the user-item similarity property, motivated by prior work~\citep{zhou-sigir23}, popular items are more broadly similar with users than long-tail items. Thus, $\bar{r}^{(l)}_{p,t} < \bar{r}^{(l)}_{q,t}$ with $\bar{r}^{(l)}_{q,t} > 0$, which implies
$1 - \bar{r}^{(l)}_{p,t} / \bar{r}^{(l)}_{q,t} > 0$. Combined with the positivity of
$\mathcal{I}^{(l)}_p / \mathcal{I}^{(l)}_q$, the product is strictly positive, establishing both the equality and the positivity in Lemma~\ref{lemma:IIW_reduction}.

Lemma~\ref{lemma:IIW_reduction} has two implications for understanding IIW. First, the larger the popularity gap between popular and long-tail items, reflected by a smaller IIW ratio $\bar{r}^{(l)}_{p,t} / \bar{r}^{(l)}_{q,t}$, the larger the IIW-induced reduction in popular-item dominance. Second, the larger the structural degree imbalance $d_p / d_q$, the larger the reduction in the relative contribution ratio, because the standard contribution ratio
$\mathcal{I}^{(l)}_p / \mathcal{I}^{(l)}_q$ scales linearly with the degree ratio. Therefore, IIW corrects more strongly when popularity bias is more pronounced, either in the learned embeddings or in the graph topology.

\subsubsection{Proof of Lemma~\ref{lemma:lw}}
\label{app:lw-analysis}

We analyze how the LW component of DPAA shapes representation learning.
To understand the effect of $\lambda_\ell$ on the embeddings, we analyze how it affects the gradient of the loss function during backpropagation. The BPR loss is the standard training objective in LightGCN, and the gradient of this loss with respect to a node's embedding directly determines how that embedding evolves during training. By tracing how $\lambda_\ell$ propagates through this gradient, we can understand how LW reshapes the contribution of different layers to representation updates. We perform this analysis using a simplified LightGCN-style propagation for simplicity. For notational simplicity, we omit the LightGCN degree-normalization constants since including them does not change the role of $\lambda_\ell$ in the gradient expression.

For a training triple $(u,i,j)$, where $u$ is a user, $i$ is an item with which $u$ has interacted, and $j$ is a randomly sampled item with which $u$ has not interacted, the BPR loss is:
\begin{equation*}
    \mathcal{L}_{uij}
    =
    -\ln \sigma(\hat{y}_{ui}-\hat{y}_{uj}),
\end{equation*}
where
$
    \hat{y}_{ui}
    =
    \mathbf{e}_u^\top \mathbf{e}_i$,  
    $\hat{y}_{uj}
    =
    \mathbf{e}_u^\top \mathbf{e}_j$,
and $\mathbf{e}_u$, $\mathbf{e}_i$, and $\mathbf{e}_j$ are the final embeddings.

Let
$
    s_{uij}
    =
    \hat{y}_{ui}-\hat{y}_{uj}
    =
    \mathbf{e}_u^\top(\mathbf{e}_i-\mathbf{e}_j).
$
Differentiating the BPR loss with respect to $s_{uij}$ gives:
\begin{equation*}
    \frac{\partial \mathcal{L}_{uij}}{\partial s_{uij}}
    =
    -\epsilon_{uij},
    \qquad
    \epsilon_{uij}
    =
    1-\sigma(s_{uij}).
\end{equation*}
By the chain rule, for any node $v\in\{u,i,j\}$:
\begin{equation} \label{eq:loss_grad_with_node}
    \frac{\partial \mathcal{L}_{uij}}{\partial \mathbf{e}_v}
    =
    -\epsilon_{uij}
    \cdot
    \frac{\partial s_{uij}}{\partial \mathbf{e}_v}.
\end{equation}
The gradients of $s_{uij}$ with respect to node embeddings are:
\begin{equation} \label{eq:s_with_node_grad}
    \frac{\partial s_{uij}}{\partial \mathbf{e}_u}
    =
    \mathbf{e}_i-\mathbf{e}_j,
    \qquad
    \frac{\partial s_{uij}}{\partial \mathbf{e}_i}
    =
    \mathbf{e}_u,
    \qquad
    \frac{\partial s_{uij}}{\partial \mathbf{e}_j}
    =
    -\mathbf{e}_u .
\end{equation}

For the gradient analysis, we first express the final embedding of a node $v$ using the LightGCN readout function in terms of its layer-wise representations as:
\begin{equation*}
    \mathbf{e}_v
    =
    \frac{1}{L+1}
    \sum_{\ell=0}^{L}
    \mathbf{e}_v^{(\ell)} .
\end{equation*}
Thus, for item $i$:
\begin{equation*}
    \mathbf{e}_i
    =
    \frac{1}{L+1}
    \sum_{\ell=0}^{L}
    \mathbf{e}_i^{(\ell)} .
\end{equation*}
Substituting the LightGCN-style message passing with LW:
$
    \mathbf{e}_i^{(\ell)}
    =
    \lambda_{\ell-1} \sum_{u'\in\mathcal{N}(i)}
    \mathbf{e}_{u'}^{(\ell-1)},
$
using $\lambda_0=1$ and $\lambda_{\ell-1}=\ell^\eta$, we obtain:
\begin{equation*}
    \mathbf{e}_i
    =
    \frac{1}{L+1}
    \left(
        \mathbf{e}_i^{(0)}
        +
        \sum_{\ell=1}^{L}
        \ell^\eta
        \sum_{u'\in\mathcal{N}(i)}
        \mathbf{e}_{u'}^{(\ell-1)}
    \right).
\end{equation*}
The same expansion applies to $\mathbf{e}_j$ and $\mathbf{e}_u$:
\begin{equation*}
    \mathbf{e}_j
    =
    \frac{1}{L+1}
    \left(
        \mathbf{e}_j^{(0)}
        +
        \sum_{\ell=1}^{L}
        \ell^\eta
        \sum_{u'\in\mathcal{N}(j)}
        \mathbf{e}_{u'}^{(\ell-1)}
    \right),
\end{equation*}
and
\begin{equation*}
    \mathbf{e}_u
    =
    \frac{1}{L+1}
    \left(
        \mathbf{e}_u^{(0)}
        +
        \sum_{\ell=1}^{L}
        \ell^\eta
        \sum_{i'\in\mathcal{N}(u)}
        \mathbf{e}_{i'}^{(\ell-1)}
    \right).
\end{equation*}

Substituting these expansions into \eqref{eq:loss_grad_with_node},  using values from \eqref{eq:s_with_node_grad}, and combining the three cases gives the gradient with respect to any node $v\in\{u,i,j\}$:
{\scriptsize
\begin{equation*}
\frac{\partial \mathcal{L}_{uij}}{\partial \mathbf{e}_v}
=
\begin{cases}
\displaystyle
\frac{-\epsilon_{uij}}{L+1}
\left(
    \mathbf{e}_i^{(0)}-\mathbf{e}_j^{(0)}
    +
    \sum_{\ell=1}^{L}
    \ell^\eta
    \left[
        \sum_{u'\in\mathcal{N}(i)}
        \mathbf{e}_{u'}^{(\ell-1)}
        -
        \sum_{u'\in\mathcal{N}(j)}
        \mathbf{e}_{u'}^{(\ell-1)}
    \right]
\right),
& \text{if } v=u,
\\[10pt]

\displaystyle
\frac{-\epsilon_{uij}}{L+1}
\left(
    \mathbf{e}_u^{(0)}
    +
    \sum_{\ell=1}^{L}
    \ell^\eta
    \sum_{i'\in\mathcal{N}(u)}
    \mathbf{e}_{i'}^{(\ell-1)}
\right),
& \text{if } v=i,
\\[10pt]

\displaystyle
\frac{\epsilon_{uij}}{L+1}
\left(
    \mathbf{e}_u^{(0)}
    +
    \sum_{\ell=1}^{L}
    \ell^\eta
    \sum_{i'\in\mathcal{N}(u)}
    \mathbf{e}_{i'}^{(\ell-1)}
\right),
& \text{if } v=j .
\end{cases}
\end{equation*}
}

In each case, since $\frac{\epsilon_{uij}}{L+1}$ is a scalar constant with respect 
to the embeddings, the gradient is proportional to the remaining term, which can be 
written as:
\begin{equation*}
    \frac{\partial \mathcal{L}_{uij}}{\partial \mathbf{e}_v} 
    \propto 
    \phi_v(\mathbf{e}^{(0)}) 
    + 
    \sum_{\ell=1}^{L} \ell^\eta \cdot \psi_\ell(\mathbf{h}_v^{(\ell)})
\end{equation*}
where $\phi_v(\mathbf{e}^{(0)})$ is a function of the initial embeddings, 
and $\psi_\ell(\mathbf{h}_v^{(\ell)})$ is a function of the layer-$\ell$ 
neighborhood aggregation of node $v$. In each gradient expression, the contribution from a layer embeddings carries a multiplicative factor of $\ell^\eta$. The downstream effect on representation learning is the following.

In each gradient expression, the contribution from a layer-$\ell$ embedding 
carries a multiplicative factor of $\ell^\eta$, which directly determines how strongly that layer participates in shaping the representation update. The downstream effect on representation learning can be characterized along two complementary directions.

\textit{First, structural amplification of deeper layers.} For any two layers $\ell_a$ and $\ell_b$ with $\ell_a < \ell_b$, the deeper layer carries a strictly larger gradient coefficient $(\ell_a)^\eta < (\ell_b)^\eta$ whenever $\eta > 0$. In contrast, a standard GNN without LW corresponds to the special case $\eta = 0$, where $\ell^\eta = 1$ for every layer and all layers contribute equally to the gradient, regardless of their depth. By setting $\eta > 0$, LW therefore structurally amplifies the gradient influence of deeper layers during backpropagation, reducing the relative contribution of shallow layers whose updates are driven primarily by immediate user--item interactions. Since deeper layers aggregate signals from multi-hop neighborhoods, this re-weighting distributes representation learning across a broader set of items reachable through indirect collaborative paths, including diverse and potentially long-tail items that lie beyond a user's immediate neighborhood.

\textit{Second, monotonic strengthening with $\eta$.} As $\eta$ increases, the gradient-coefficient ratio $(\ell_b)^\eta / (\ell_a)^\eta > 1$ between any deeper layer $\ell_b$ and any shallower layer $\ell_a$ grows monotonically, progressively amplifying the contribution of deeper layers to embedding updates. This provides a mechanism for controlling how strongly higher-order collaborative signals influence representation learning relative to shallow ones. In scenarios where immediate interactions are heavily dominated by popular items, as is typical under severe popularity skew, a larger $\eta$ can be particularly beneficial, since it shifts gradient influence away from popularity-skewed local interactions and toward higher-order collaborative signals carried through multi-hop neighborhoods. Conversely, when popularity bias is mild, smaller values of $\eta$ can suffice, as shallow-layer updates are not as heavily distorted by popularity-driven interactions.

Together, these two implications from Lemma~\ref{lemma:lw} explain how LW reshapes representation learning: it not only amplifies gradient influence toward deeper layers in a structural sense, but also provides a tunable knob ($\eta$) that scales this amplification with the severity of local dominance of popularity bias in the data, consistent with the motivation for incorporating LW into message passing.

\subsection{Additional Experimental Details}
\label{app:experimental_detail}

\subsubsection{Dataset details} \label{app:dataset_deatils}
We use three real-world datasets with unbiased test sets: Coat~\citep{schnabel-icml16}, KuaiRec~\citep{gao-cikm22}, and Yahoo! R3~\citep{marlin-recsys09}. Coat contains user ratings from an online coat shopping platform. Coat contains 290 users, 300 items, and 5,490 interactions, with a sparsity of 0.06310. In the training set, each user rates 24 self-selected items, while the test set is constructed by randomly exposing 16 items to each user. Ratings of 3 or higher on the 5-point scale are treated as positive feedback. KuaiRec is derived from Kuaishou short-video logs and provides a fully observed user--item matrix for evaluation,  where nearly all users interact with almost every video. KuaiRec contains 7,176 users, 10,728 items, and 1,153,106 interactions overall, with a sparsity of 0.01498. Although this test setting is dense, the training data remains sparse. Its test set contains ratings from 1,411 users over 3,327 items. Following~\citep{gao-cikm22}, we treat an interaction as positive if the watch time exceeds twice the video duration. Yahoo! R3 includes user--song ratings collected under self-selection. Yahoo! R3 contains 14,382 users, 1,000 items, and 129,748 interactions, with a sparsity of 0.00902. Yahoo! R3 is substantially sparser than Coat and KuaiRec, allowing us to evaluate model performance under a more challenging sparse-feedback setting. Its unbiased test set asks 5,400 users to rate 10 randomly selected songs. Ratings of 3 or higher are treated as positive feedback. We select these datasets because their unbiased test sets enable unbiased evaluation of recommendation methods under real-world settings. Moreover, their differences in scale, sparsity, and popularity distribution allow us to assess model performance across diverse recommendation scenarios. Dataset statistics are summarized in Table~\ref{tab:dataset_stats}.
\begin{table} [htpb] \label{tab:dataset_stats}
    \fontsize{9}{7}\selectfont
    \setlength{\tabcolsep}{6.0pt} 
    \centering
    \captionsetup{justification=raggedright, width=1.0\textwidth}
    \caption{Dataset statistics.}
    \label{tab:dataset_stats}
    \begin{tabular}{lcccc}
        \toprule
        Dataset & \#Users & \#Items & \#Interactions & Sparsity \\
        \midrule
        Coat       & 290    & 300    & 5,490     & 0.06310 \\
        KuaiRec    & 7,176  & 10,728 & 1,153,106 & 0.01498 \\
        Yahoo! R3  & 14,382 & 1,000  & 129,748   & 0.00902 \\
        \bottomrule
    \end{tabular}
\end{table}

\subsubsection{Popularity Distribution of the Datasets} \label{sec:popularity_distributions}
To examine the popularity imbalance in the datasets used in our experiments, we analyze how training interactions are distributed across popular and niche items. This analysis illustrates the extent to which interactions are concentrated among popular items in each dataset. 
\begin{wrapfigure}{r}{0.47\columnwidth}
    \centering
    \includegraphics[width=0.47\columnwidth]{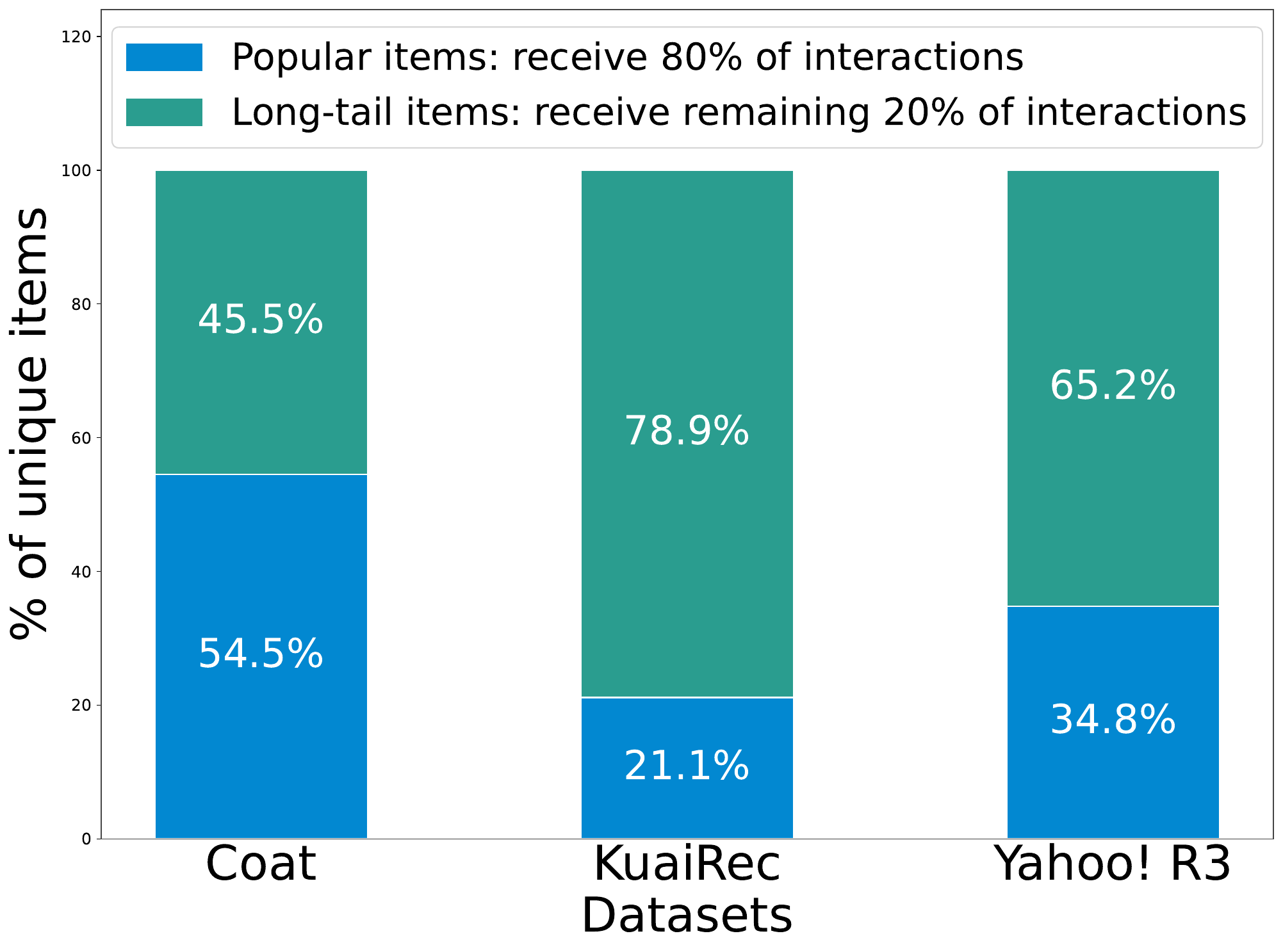}
    \captionsetup{justification=raggedright, margin=0cm}
    \captionsetup{width=0.47\textwidth}
    \caption{Training data distribution across real-world datasets. Popular items account for ${80\%}$ of interactions, while the remaining items form the long-tail group.}
    \label{fig:dataset_distribution}
\end{wrapfigure}
Figure~\ref{fig:dataset_distribution} shows the training data distributions for the three real-world datasets used in our experiments. For each dataset, popular items are defined as the smallest set of items that accounts for $80\%$ of observed training interactions, while the remaining items are treated as niche or long-tail items and account for the remaining $20\%$. On Coat, $54.5\%$ of items account for $80\%$ of training interactions, compared with only $21.1\%$ on KuaiRec and $34.8\%$ on Yahoo! R3. These proportions reveal differences in the degree of popularity imbalance across the datasets. KuaiRec exhibits the strongest popularity skew, with most interactions concentrated among a relatively small fraction of items. Yahoo! R3 shows moderate skew, while Coat exhibits the mildest skew among the three datasets, with interactions spread across a larger proportion of items. Overall, these results show that interaction distributions in these real-world datasets are skewed toward popular items, although the extent of this skew varies across datasets.

\subsubsection{Relationship between user--item similarity and item popularity}\label{sec:popularity_similarity}

\begin{figure*}[htpb]
\centering

\begin{minipage}[t]{0.50\textwidth}
    \centering
    \vspace{-4pt}
    \includegraphics[width=0.90\textwidth]{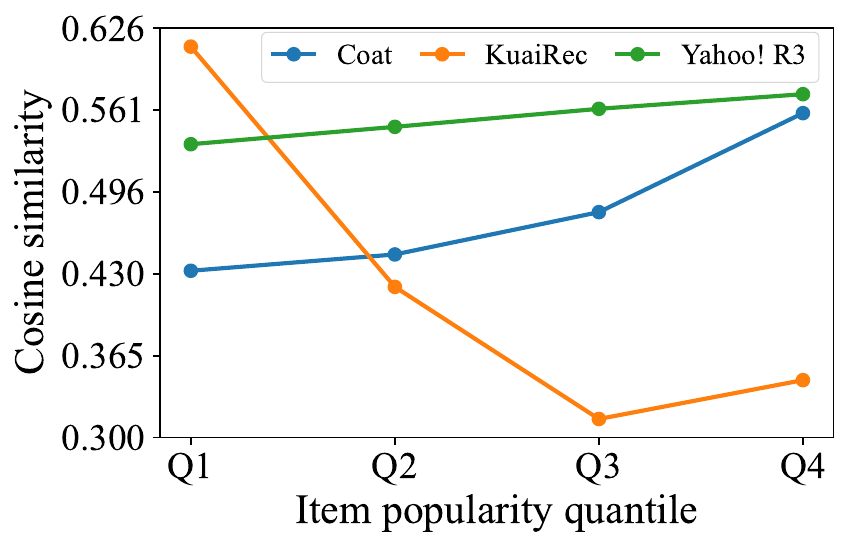}
    
    \captionsetup{justification=raggedright, width=1.0\textwidth}
    \caption{Average user--item cosine similarity across item popularity quartiles. Items are divided into four equally sized groups based on interaction frequency: Q1 contains the least popular 25\% of items, Q2 the 26--50\% range, Q3 the 51--75\% range, and Q4 the rest most popular 25\% items.}
    \label{fig:mean_cosine_similarity_quartiles}
\end{minipage}
\hfill
\begin{minipage}[t]{0.47\textwidth}
    \centering
    \vspace{0pt}
    \fontsize{8}{9}\selectfont
    \setlength{\tabcolsep}{4.0pt}
    \captionsetup{justification=raggedright, width=1.0\textwidth}
    \captionof{table}{Percentage of training interactions across item popularity quartiles. Each value  indicates the share of training interactions associated with each item quartile.}
    \label{tab:interaction_percentage_quartiles}

    \begin{tabular}{lccc}
        \toprule
        Quartile & Coat & KuaiRec & Yahoo! R3 \\
        \midrule
        Q1 (1-25\%) & 7.4\%  & 0.4\%  & 4.0\% \\
        Q2 (26-50\%) & 15.8\% & 2.0\%  & 8.2\% \\
        Q3 (51-75\%) & 24.5\% & 12.3\% & 14.7\% \\
        Q4 (76-100\%) & 52.3\% & 85.3\% & 73.0\% \\
        \bottomrule
    \end{tabular}

\end{minipage}
\end{figure*}

We empirically examine the relationship between item popularity and user--item similarity, which motivates our interaction-level weighting mechanism, where interactions with higher user--item similarity are more likely to involve popular items. We rank items by their training interaction frequency and divide them into four equally sized popularity quartiles: Q1 contains the bottom 25\% (least popular items), Q2 the 25--50\% range, Q3 the 50--75\% range, and Q4 the 75--100\% range (most popular 25\%). For each quartile, we compute the average cosine similarity over observed user--item interactions using the pre-trained model, as shown in Figure\ref{fig:mean_cosine_similarity_quartiles}, and also report the percentage of training interactions belonging to each quartile alongside it in Table~\ref{tab:interaction_percentage_quartiles}. Coat dataset exhibits a clear increasing trend, with the average cosine similarity increasing from Q1 to Q4. Yahoo! R3 shows a similar overall pattern, with higher average similarity for more popular items. KuaiRec shows a different distribution: the first two popularity quartiles (Q1 and Q2) account for only 2.4\% of the observed interactions, whereas Q3 and Q4 together account for approximately 97.6\%. Although Q1 and Q2 show higher similarity than Q3, they account for only a small fraction of interactions, and therefore have little influence on the overall interaction distribution. Thus, Q3 and Q4 are representative of the interaction distribution in KuaiRec. Within these two dominant groups, the average cosine similarity increases from Q3 to Q4 as item popularity increases. Overall, these results provide empirical evidence that interactions involving more popular items generally tend to exhibit higher user--item similarity. This supports our use of the complement of cosine similarity as an inverse interaction weight during aggregation, assigning smaller weights to interactions involving popular items and larger weights to those involving niche items.

\subsubsection{Unbiased evaluations and ranking protocol} \label{app:unbiased_eval}
We choose the above mentioned three real-world datasets because they provide unbiased test set for unbiased evaluations. While other datasets could also be used for experiments, they typically lack unbiased test sets, as their test data is randomly sampled from observed interactions in the training data~\citep{wei-sigkdd21, zhou-sigir23}. Even when randomly sampled, such test data can still reflect the underlying popularity bias present in the observed interactions. This is because the samples are drawn from popularity-dominated distributions, which limits their suitability for unbiased evaluation. In contrast, the chosen datasets offer unbiased test sets via random exposure or full observation, and span diverse sizes, sparsity levels, and popularity distributions in the training data, reflecting different real-world scenarios.

We follow the all-ranking protocol~\citep{krichene-sigkdd20}, ranking all candidate items for each user after removing positive training interactions.
We report Recall@$k$, NDCG@$k$, and Hit Ratio (HR@$k$), with $k=20$ by default.
For KuaiRec, we follow the modified protocol as in~\citep{zhang-neurips22}, since its test set comes from a densely observed sub-matrix rather than a random sample over the full item catalog~\citep{gao-cikm22}. Thus, following ~\citep{zhang-neurips22}, ranking is restricted to the $3{,}327$ fully exposed items in the test sub-matrix.

\subsubsection{Details of semi-synthetic click data generation process}
\label{app:syn_data_gen_details}
To systematically evaluate how debiasing methods perform under different levels of popularity bias, we construct semi-synthetic training data with controllable bias severity for RQ2. Real-world recommendation datasets typically do not provide direct control over the degree of popularity bias in the training data, making it difficult to analyze method robustness under varying bias intensities. To address this limitation, we generate user--item interactions using a Zipfian probability sampling mechanism~\cite{newman-physics05} based on item interaction frequencies, which allows us to continuously control the skewness of the training distribution.

We use the original unbiased KuaiRec test set as the basis for dataset construction because it is collected under full-exposure settings and contains many interactions per user. We split this data into three disjoint subsets: $10\%$ for validation, $20\%$ for testing, and the remaining $70\%$ as an unbiased interaction pool for generating popularity-skewed training data. This design ensures that popularity bias is introduced only in the training set, while validation and test sets remain unbiased for fair evaluation.

Before applying probabilistic sampling, we enforce a minimum exposure constraint to avoid degenerate cases. Specifically, for each item, we randomly select one user from the set of users who originally interacted with that item and add the corresponding interaction to the training data, ensuring that every item appears at least once. We then compute a global popularity ranking by counting item frequencies and sorting items in descending order. Let $r$ denote the popularity rank of an item and $N$ denote the total number of items. The Zipfian sampling probability for an item with rank $r$ is defined as:
\begin{equation*}
P_{\text{zipf}}(r) = \frac{1 / r^{s}}{\sum_{n=1}^{N} 1 / n^{s}},
\end{equation*}
where $s$ controls the skewness of the distribution. Smaller values of $s$ produce near-uniform sampling, while larger values concentrate more probability on top-ranked popular items. We therefore refer to $s$ as the popularity bias severity parameter. For each user, we normalize these Zipfian probabilities over the user's candidate item set and sample interactions accordingly. This preserves each user's original item pool while injecting global popularity bias into the generated training interactions. By varying $s$ from $0$ to $9$, we obtain a set of datasets ranging from weakly biased to highly popularity-skewed, enabling systematic evaluation of debiasing methods under increasing popularity bias.

\subsubsection{Baseline details}
\label{app:baselines}
Our setup is based on pure CF setting. We use only user-item interactions without any auxiliary item features or user attributes. Therefore, we compare DPAA against baselines that are compatible with this setting. Detailed descriptions of baselines are provided below.
\begin{itemize}[leftmargin=9pt, nosep]
    \item \textbf{APDA}~\citep{zhou-sigir23} mitigates aggregation-stage bias in GNN-based CF by assigning each edge an adaptive weight derived from an inverse popularity score during message passing.
    \item \textbf{NAVIP}~\citep{kim-cikm22} debiases neighborhood aggregation in GNNs by using inverse propensity scores as edge weights, reducing the influence of popular items during message passing.
    \item \textbf{DAP}~\citep{chen-front24} is a post-hoc method that clusters structurally similar nodes, estimates the popularity-amplification component of their embeddings, and subtracts it to obtain unbiased representations.
    \item \textbf{PPAC}~\citep{ning-www24} jointly models personal and global popularity within a counterfactual inference framework to attribute and remove popularity-driven prediction effects.
    \item \textbf{MACR}~\citep{wei-sigkdd21} applies counterfactual inference within a multi-task learning framework to isolate and subtract the direct causal effect of item popularity on the predicted score.
    \item \textbf{CVIB}~\citep{wang-neurips20} applies an 
    information-theoretic counterfactual method to learn unbiased recommendations from missing-not-at-random data by minimizing mutual information to correct for bias.
    \item \textbf{IPS}~\citep{gruson-wsdm19} extends the standard IPW estimator~\citep{joachims-wsdm17} with max-capping and normalization of the propensity weights, which reduces estimator variance.
    \item \textbf{SAM-REG}~\citep{boratto-ipm21} pairs balanced sampling across popularity tiers with a regularizer that penalizes the correlation between predicted relevance and item popularity.
    \item \textbf{LightGCN}~\citep{he-sigir20} is a GNN-based CF backbone without bias correction.
\end{itemize}

\subsubsection{Implementation details} \label{app:impl_details}
Unless stated otherwise, we use LightGCN~\citep{he-sigir20} as the backbone for all baselines and our method, since it is one of the most widely used architectures for GNN-based CF~\citep{zhou-sigir23}. We use the LightGCN implementation of PyTorch and train it with the Adam optimizer. We apply early stopping with a maximum of $1000$ epochs and a patience of 50 epochs. Training is terminated if Recall@$20$ on the validation set does not improve for 50 consecutive epochs, ensuring good performance under the early stopping criterion. We tune the learning rate over ${1{e}^{-5}, 1{e}^{-4}, 1{e}^{-3}}$. We set the batch size to 2,048 and the embedding dimension to 256, while keeping all other parameters at the default settings of LightGCN. We use a two-layer GCN for the Coat dataset and Yahoo! R3 dataset, and a four-layer GCN for the KuaiRec dataset. Some initial analysis on LightGCN shows that deeper architectures perform better on KuaiRec, whereas fewer layers yield better results on Coat and Yahoo! R3. We therefore apply this same configuration consistently across our method and all baselines for fair comparison. We tune the epoch-dependent coefficient for transitioning from pre-trained to current embeddings over $C \in \{0.0, 1{e}^{-5}, 1{e}^{-4}, 1{e}^{-3}, 1{e}^{-2}, 1{e}^{-1}, 1{e}^{0}\}$. We tune the layer-wise weighting parameter, which controls the relative importance of different layers, over $\eta \in \{0, 1, 2, 3\}$. We also tune the initial residual connection parameter $\delta$, which balances initial embeddings and neighborhood aggregation, over the range $[0.0, 1.0]$ with a step size of 0.2. Unless stated otherwise, for the pre-trained model used to estimate IIW, we use a basic LightGCN model without any bias correction.

For RQ4 only, we use the more recent LightGCN++~\citep{lee-tors26} as the backbone to evaluate the generalizability of DPAA to a different GNN-based CF architecture and its ability to mitigate popularity bias in that backbone. In this setting, we also use LightGCN++ as the pre-trained model for estimating IIW. LightGCN++ extends LightGCN by introducing flexible embedding-norm scaling and neighbor weighting during aggregation, together with an adjustable layer-wise embedding pooling mechanism to address the inflexibility and inconsistency of the original LightGCN architecture~\citep{lee-tors26}. For LightGCN++, we use the implementation provided by the authors and follow the parameter settings of the original LightGCN++ implementation.

Unless stated otherwise, in \eqref{eq:final_agg_weight}, we set $\gamma=1$ for all experiments, meaning that IIW is applied only at the initial layer. 
In our ablation studies, we also provide an analysis using $\gamma=0$, meaning that IIW is applied at all layers in that case. We conduct our experiments on machines with 252 GB RAM, 3 TB storage, a CUDA-enabled GPU with 24 GB memory, 128 CPU cores, and a clock speed of 1.46 GHz.

\subsection{Additional Experimental Results} \label{app:add_results}
\subsubsection{More analysis on RQ1} \label{app:add_results_rq1}
We now analyze the different parameters of DPAA that produce the best results across datasets and discuss the role of each parameter in adapting our method to different data characteristics.  Table~\ref{tab:dpaa_parameter_combination} summarizes the best parameter combinations of DPAA across datasets: $C$, which controls the transition from pre-trained to current model embeddings, $\eta$, which controls layer-wise weighting, and $\delta$, which controls the initial residual connection.

On Coat, the best parameter combination uses $C=1e^{-5}$, $\eta=2.0$, and $\delta=0.2$. The small but nonzero value of $C$ allows a gradual transition from pre-trained embeddings to current model embeddings, suggesting that current embeddings can also provide useful adaptive signals once training becomes stable. The moderate value of $\eta$ indicates that higher-order collaborative signals are useful even under milder popularity bias. The small value of $\delta$ suggests that preserving a limited amount of initial embedding information is sufficient to reduce over-smoothing while still allowing neighborhood aggregation to shape the representations.

On KuaiRec, the best combination uses $C=1e^{-3}$, $\eta=3.0$, and $\delta=0.2$. The comparatively moderate value of $C$ allows the model to rely on both pre-trained and current model embeddings during training. The larger $\eta$ is also consistent with the stronger popularity skew in KuaiRec, where emphasizing deeper-layer collaborative signals helps reduce the dominance of shallow popularity-driven interactions, consistent with Lemma~\ref{lemma:lw}. The small value of $\delta$ suggests that preserving a limited amount of initial embedding information is sufficient to reduce over-smoothing.

On Yahoo! R3, the best combination uses $C=0.0$, $\eta=0.0$, and $\delta=0.6$. Since $C=0.0$ means the model does not use current model embeddings for IIW estimation and relies only on the pre-trained embedding-based weights, this suggests that current embeddings may be unreliable in sparse scenarios, showing the importance of using the pre-trained model. The value $\eta=0.0$ indicates that uniform layer contribution works better than emphasizing deeper layers. The moderate value of $\delta$ suggests that preserving some initial embedding information is necessary while still allowing the model to incorporate higher-order neighbors.

\begin{table} [htpb]
    \centering
    \fontsize{9}{7}\selectfont
    \setlength{\tabcolsep}{6.0pt} 
    \caption{Best parameter combinations of DPAA across datasets. Here, $C$ controls the transition from pre-trained to current model embeddings, $\eta$ controls layer-wise weighting, and $\delta$ controls the initial residual connection.}
    \label{tab:dpaa_parameter_combination}
    \begin{tabular}{lccc}
        \toprule
        Dataset & $C$ & $\eta$ & $\delta$ \\
        \midrule
        Coat & $1e^{-5}$ & 2.0 & 0.2 \\
        KuaiRec & $1e^{-3}$ & 3.0 & 0.2 \\
        Yahoo! R3 & 0.0 & 0.0 & 0.6 \\
        \bottomrule
    \end{tabular}
\end{table}

Overall, these dataset-specific combinations show that DPAA adapts to different recommendation settings and highlight the importance of its key components, including the combination of pre-trained and current model estimates, the gradual transition between them for IIW estimation, layer-wise weighting, and the preservation of initial embedding information to reduce over-smoothing.

\subsubsection{More analysis on RQ2} \label{app:add_results_rq2}
In the main body of the paper, we present the Recall@20 results under varying popularity bias severity in Figure~\ref{fig:severity} on the KuaiRec semi-synthetic dataset. For better interpretation of the exact value under different severity levels s, we also report the corresponding Recall@20 results in Table~\ref{tab:main_results_unbiased_synthetic}.

\begin{table*} [htpb]
\centering
\fontsize{6.0}{8}\selectfont
\setlength{\tabcolsep}{2.0pt}
\newcommand{\stdsub}[1]{$_{\scalebox{0.7}{$\scriptscriptstyle\pm#1$}}$}
\caption{Performance comparison of different methods for Recall@20 on the KuaiRec dataset by varying levels of popularity bias severity ($\bm{s}$). Bold and underline indicate the best and second-best results. Results are reported as mean $\bm{\pm}$ std\%. Improvement (\%) denotes the relative gain of DPAA over the best baseline.}
\begin{tabular}{cccccccccccc}
\toprule
$s$ & DPAA & APDA & NAVIP & DAP & PPAC & MACR & CVIB & IPS & SAM-REG & LightGCN & Improvement (\%) \\
\midrule
0.0  & \textbf{0.2047}\stdsub{0.70} & 0.1740\stdsub{0.46} & 0.1557\stdsub{0.18} & 0.1738\stdsub{0.83} & \underline{0.2025}\stdsub{0.31} & 0.1213\stdsub{0.57} & 0.1127\stdsub{0.09} & 0.0987\stdsub{0.74} & 0.0956\stdsub{0.26} & 0.2024\stdsub{0.91} & +1.09 \\
1.0  & \underline{0.2043}\stdsub{1.11} & 0.2001\stdsub{0.72} & 0.1941\stdsub{0.34} & \textbf{0.2049}\stdsub{0.96} & 0.1949\stdsub{1.12} & 0.0989\stdsub{1.21} & 0.1762\stdsub{1.24} & 0.1691\stdsub{0.48} & 0.1597\stdsub{0.87} & 0.2001\stdsub{0.29} & -0.29 \\
2.0  & \textbf{0.1859}\stdsub{1.01} & 0.1763\stdsub{0.58} & 0.1613\stdsub{0.75} & 0.1769\stdsub{0.92} & 0.1810\stdsub{0.43} & 0.1186\stdsub{0.76} & 0.1663\stdsub{0.21} & 0.1210\stdsub{0.64} & 0.1722\stdsub{0.35} & \underline{0.1824}\stdsub{0.88} & +1.92 \\
3.0  & \textbf{0.1703}\stdsub{0.53} & 0.1453\stdsub{0.29} & 0.0920\stdsub{0.67} & 0.1325\stdsub{0.41} & 0.1550\stdsub{0.82} & 0.0944\stdsub{0.55} & 0.1373\stdsub{0.32} & 0.0509\stdsub{1.14} & 0.1600\stdsub{0.24} & \underline{0.1656}\stdsub{0.61} & +2.84 \\
4.0  & \textbf{0.1500}\stdsub{1.22} & 0.1146\stdsub{0.81} & 0.0574\stdsub{1.21} & 0.0798\stdsub{1.06} & 0.1017\stdsub{0.52} & 0.0037\stdsub{0.74} & 0.0558\stdsub{1.31} & 0.0314\stdsub{0.43} & 0.0461\stdsub{0.96} & \underline{0.1215}\stdsub{0.68} & +23.46 \\
5.0  & \textbf{0.1210}\stdsub{1.01} & \underline{0.1148}\stdsub{0.63} & 0.0367\stdsub{0.24} & 0.0497\stdsub{0.91} & 0.0759\stdsub{0.47} & 0.0083\stdsub{0.72} & 0.0324\stdsub{0.35} & 0.0226\stdsub{0.56} & 0.0211\stdsub{0.82} & 0.0804\stdsub{0.39} & +5.40 \\
6.0  & \textbf{0.1008}\stdsub{0.81} & \underline{0.0853}\stdsub{0.42} & 0.0317\stdsub{0.69} & 0.0389\stdsub{0.33} & 0.0567\stdsub{0.94} & 0.0083\stdsub{0.57} & 0.0226\stdsub{0.26} & 0.0234\stdsub{0.73} & 0.0179\stdsub{1.18} & 0.0477\stdsub{1.13} & +18.17 \\
7.0  & \textbf{0.0958}\stdsub{0.30} & \underline{0.0951}\stdsub{0.47} & 0.0274\stdsub{0.16} & 0.0298\stdsub{0.62} & 0.0510\stdsub{0.34} & 0.0080\stdsub{0.99} & 0.0179\stdsub{0.53} & 0.0190\stdsub{0.21} & 0.0114\stdsub{0.41} & 0.0400\stdsub{0.67} & +0.74 \\
8.0  & \textbf{0.0879}\stdsub{0.32} & \underline{0.0785}\stdsub{0.71} & 0.0233\stdsub{0.45} & 0.0286\stdsub{0.59} & 0.0428\stdsub{0.83} & 0.0080\stdsub{0.27} & 0.0121\stdsub{0.56} & 0.0136\stdsub{0.91} & 0.0076\stdsub{0.64} & 0.0349\stdsub{0.24} & +11.97 \\
9.0  & \textbf{0.0780}\stdsub{0.52} & \underline{0.0633}\stdsub{0.34} & 0.0186\stdsub{0.71} & 0.0216\stdsub{0.76} & 0.0373\stdsub{0.63} & 0.0073\stdsub{0.28} & 0.0098\stdsub{0.47} & 0.0111\stdsub{0.98} & 0.0078\stdsub{0.76} & 0.0278\stdsub{0.41} & +23.22 \\
\bottomrule
\end{tabular}\label{tab:main_results_unbiased_synthetic}
\end{table*}

\subsubsection{More analysis on RQ4} \label{app:add_results_rq4}
In the main body of the paper, Figure~\ref{fig:severity_lightgcn++} presents the Recall@20 results on the semi-synthetic KuaiRec dataset using LightGCN++ as the backbone under varying levels of popularity bias. To facilitate comparison across all metrics at each severity level $s$, Table~\ref{tab:synthetic_bias_dpaa_lightgcn++} reports the corresponding Recall@20, NDCG@20, and HR@20 results. Consistent with the Recall@20 findings discussed earlier, DPAA outperforms the baselines across all three metrics under both mild and severe bias.

\begin{table*}[hbtp]
\fontsize{5.8}{8}\selectfont
\setlength{\tabcolsep}{1.8pt}
\providecommand{\stdsub}[1]{$_{\scalebox{0.7}{$\scriptscriptstyle\pm#1$}}$}
\captionsetup{justification=raggedright, width=1.0\textwidth}
\caption{Performance comparison of different methods using LightGCN++ as the backbone on the semi-synthetic KuaiRec dataset under varying levels of popularity bias severity ($s$). We include the LightGCN++ and APDA, the best-performing baseline, for comparison. Bold and underline indicate the best and second-best results. Results are reported as mean $\bm{\pm}$ std\%. Improvement (\%) denotes the relative gain of DPAA over the best performing method.}
\centering

\begin{tabular}{llcccccccccc}
\toprule
\multirow{2}{*}{Metric} &
\multirow{2}{*}{Method} &
\multicolumn{10}{c}{Popularity Bias Severity ($s$)} \\
\cmidrule(lr){3-12}
& & 0 & 1 & 2 & 3 & 4 & 5 & 6 & 7 & 8 & 9 \\
\midrule

\multirow{3}{*}{Recall}
& DPAA (Ours)
& \underline{0.1953}\stdsub{0.79}
& \textbf{0.1999}\stdsub{0.68}
& \textbf{0.1814}\stdsub{0.81}
& \textbf{0.1354}\stdsub{0.74}
& \textbf{0.1011}\stdsub{0.92}
& \textbf{0.0950}\stdsub{0.77}
& \textbf{0.0919}\stdsub{1.36}
& \textbf{0.0810}\stdsub{1.02}
& \textbf{0.0649}\stdsub{0.79}
& \textbf{0.0584}\stdsub{1.08} \\

& APDA
& 0.0286\stdsub{0.61}
& 0.0155\stdsub{0.79}
& 0.0222\stdsub{0.67}
& 0.0517\stdsub{0.91}
& 0.0705\stdsub{1.33}
& \underline{0.0799}\stdsub{1.04}
& \underline{0.0788}\stdsub{0.82}
& \underline{0.0747}\stdsub{1.11}
& \underline{0.0592}\stdsub{0.76}
& \underline{0.0519}\stdsub{1.21} \\

& LightGCN++
& \textbf{0.1954}\stdsub{0.93}
& \underline{0.1990}\stdsub{0.71}
& \underline{0.1803}\stdsub{0.84}
& \underline{0.1344}\stdsub{0.67}
& \underline{0.0929}\stdsub{1.05}
& 0.0652\stdsub{1.23}
& 0.0488\stdsub{0.89}
& 0.0401\stdsub{1.16}
& 0.0322\stdsub{0.78}
& 0.0374\stdsub{0.96} \\

\cmidrule(lr){3-12}
& Improvement (\%)
& -0.05 & +0.45 & +0.61 & +0.74 & +8.83
& +18.90 & +16.62 & +8.43 & +9.63 & +12.52 \\

\midrule

\multirow{3}{*}{NDCG}
& DPAA (Ours)
& \underline{0.1721}\stdsub{0.58}
& \textbf{0.2093}\stdsub{0.73}
& \textbf{0.1881}\stdsub{0.66}
& \textbf{0.1509}\stdsub{0.81}
& \textbf{0.1110}\stdsub{0.94}
& \textbf{0.1043}\stdsub{0.69}
& \textbf{0.0989}\stdsub{0.87}
& \textbf{0.0879}\stdsub{0.76}
& \textbf{0.0735}\stdsub{1.01}
& \textbf{0.0660}\stdsub{0.83} \\

& APDA
& 0.0359\stdsub{0.62}
& 0.0264\stdsub{0.77}
& 0.0315\stdsub{0.69}
& 0.0642\stdsub{0.88}
& 0.0811\stdsub{0.74}
& \underline{0.0887}\stdsub{1.06}
& \underline{0.0881}\stdsub{0.81}
& \underline{0.0813}\stdsub{1.12}
& \underline{0.0694}\stdsub{0.72}
& \underline{0.0597}\stdsub{0.91} \\

& LightGCN++
& \textbf{0.1728}\stdsub{0.85}
& \underline{0.2084}\stdsub{0.64}
& \underline{0.1872}\stdsub{0.79}
& \underline{0.1501}\stdsub{0.97}
& \underline{0.1104}\stdsub{0.71}
& 0.0829\stdsub{0.82}
& 0.0635\stdsub{0.93}
& 0.0514\stdsub{1.08}
& 0.0403\stdsub{0.75}
& 0.0479\stdsub{1.03} \\

\cmidrule(lr){3-12}
& Improvement (\%)
& -0.41 & +0.43 & +0.48 & +0.53 & +0.54
& +17.59 & +12.26 & +8.12 & +5.91 & +10.55 \\

\midrule

\multirow{2}{*}{HR}
& DPAA (Ours)
& \underline{0.8835}\stdsub{1.36}
& \textbf{0.9314}\stdsub{1.39}
& \textbf{0.9060}\stdsub{1.48}
& \textbf{0.8223}\stdsub{1.13}
& \textbf{0.7047}\stdsub{0.97}
& \textbf{0.6968}\stdsub{1.09}
& \textbf{0.6681}\stdsub{1.02}
& \textbf{0.6163}\stdsub{1.25}
& \textbf{0.5525}\stdsub{1.09}
& \textbf{0.5119}\stdsub{1.38} \\

& APDA
& 0.3626\stdsub{1.28}
& 0.3193\stdsub{0.84}
& 0.3506\stdsub{1.51}
& 0.5131\stdsub{1.01}
& 0.5883\stdsub{0.79}
& \underline{0.6224}\stdsub{1.32}
& \underline{0.6181}\stdsub{0.87}
& \underline{0.5926}\stdsub{1.13}
& \underline{0.5230}\stdsub{1.42}
& \underline{0.4946}\stdsub{1.46} \\

& LightGCN++
& \textbf{0.8843}\stdsub{1.41}
& \underline{0.9288}\stdsub{0.75}
& \underline{0.9013}\stdsub{0.86}
& \underline{0.8165}\stdsub{0.72}
& \underline{0.6913}\stdsub{1.07}
& 0.5656\stdsub{1.43}
& 0.4670\stdsub{1.18}
& 0.4159\stdsub{0.94}
& 0.3783\stdsub{1.06}
& 0.4131\stdsub{0.89} \\

\cmidrule(lr){3-12}
& Improvement (\%)
& -0.09 & +0.28 & +0.52 & +0.71 & +1.94
& +11.95 & +8.09 & +4.00 & +5.64 & +3.50 \\

\bottomrule
\end{tabular}

\label{tab:synthetic_bias_dpaa_lightgcn++}
\end{table*}

\subsubsection{Computational complexity analysis} \label{sec:computational_complexity}
We empirically evaluate the computational cost of DPAA by comparing it with the LightGCN backbone, since DPAA is applied on top of LightGCN. This comparison quantifies the additional training cost introduced by popularity-bias correction relative to the standard LightGCN model. Our method does not require training an additional pre-trained model, as it can reuse the backbone model from a previous training stage, such as an already deployed or base recommendation model, as the pre-trained model. Its additional overhead arises only from the weighting mechanism during message passing. IIW estimation requires computing cosine similarity for each observed interaction, which incurs constant-time computation per interaction, while LW adds only a scalar multiplication at each layer. Therefore, DPAA does not introduce any expensive auxiliary operation and retains the same overall computational order as LightGCN, with only additional constant-factor weighting operations during message passing.

For each observed user--item interaction, computing IIW using cosine similarity between $d$-dimensional embeddings requires $O(d)$. Across $|\mathcal{A}|$ interactions and $L$ layers, the total cost is $O(dL|\mathcal{A}|)$. When IIW is applied only at the initial layer, as in our default setting, this reduces to $O(d|\mathcal{A}|)$. LW requires only $O(1)$ computation per layer because it involves only scalar multiplication during aggregation. Therefore, under our default setting, the overall additional overhead per epoch is $O(d|\mathcal{A}|)$. We report the average runtime per epoch for DPAA and its LightGCN backbone in Table~\ref{tab:runtime}. On Coat, runtime increases from 1.09 to 1.25 seconds/epoch, while on Yahoo! R3 it increases from 1.36 to 3.16 seconds/epoch. On KuaiRec, runtime increases from 20.16 to 75.23 seconds/epoch. Overall, DPAA introduces a manageable increase in runtime for bias correction while maintaining the same overall computational order as the standard LightGCN backbone.

\begin{table} [htpb]
    \fontsize{9}{7}\selectfont
    \setlength{\tabcolsep}{6.0pt}
    \centering
    \captionsetup{justification=raggedright, width=1.0\textwidth}
    \caption{Average training time in seconds per epoch for LightGCN and DPAA across datasets.}
    \label{tab:runtime}
    \begin{tabular}{lcc}
        \toprule
        Dataset & LightGCN (seconds/epoch) & DPAA (seconds/epoch) \\
        \midrule
        Coat       & 1.09  & 1.25 \\
        KuaiRec    & 20.16 & 75.23 \\
        Yahoo! R3  & 1.36  & 3.16 \\
        \bottomrule
    \end{tabular}
\end{table}

\subsection{Ablation Studies} \label{app:ablations}

\subsubsection{Contribution of different components of DPAA} \label{subsec:diff_components}
\begin{wrapfigure}{r}{0.47\columnwidth}
    \centering
    \includegraphics[width=0.47\columnwidth]{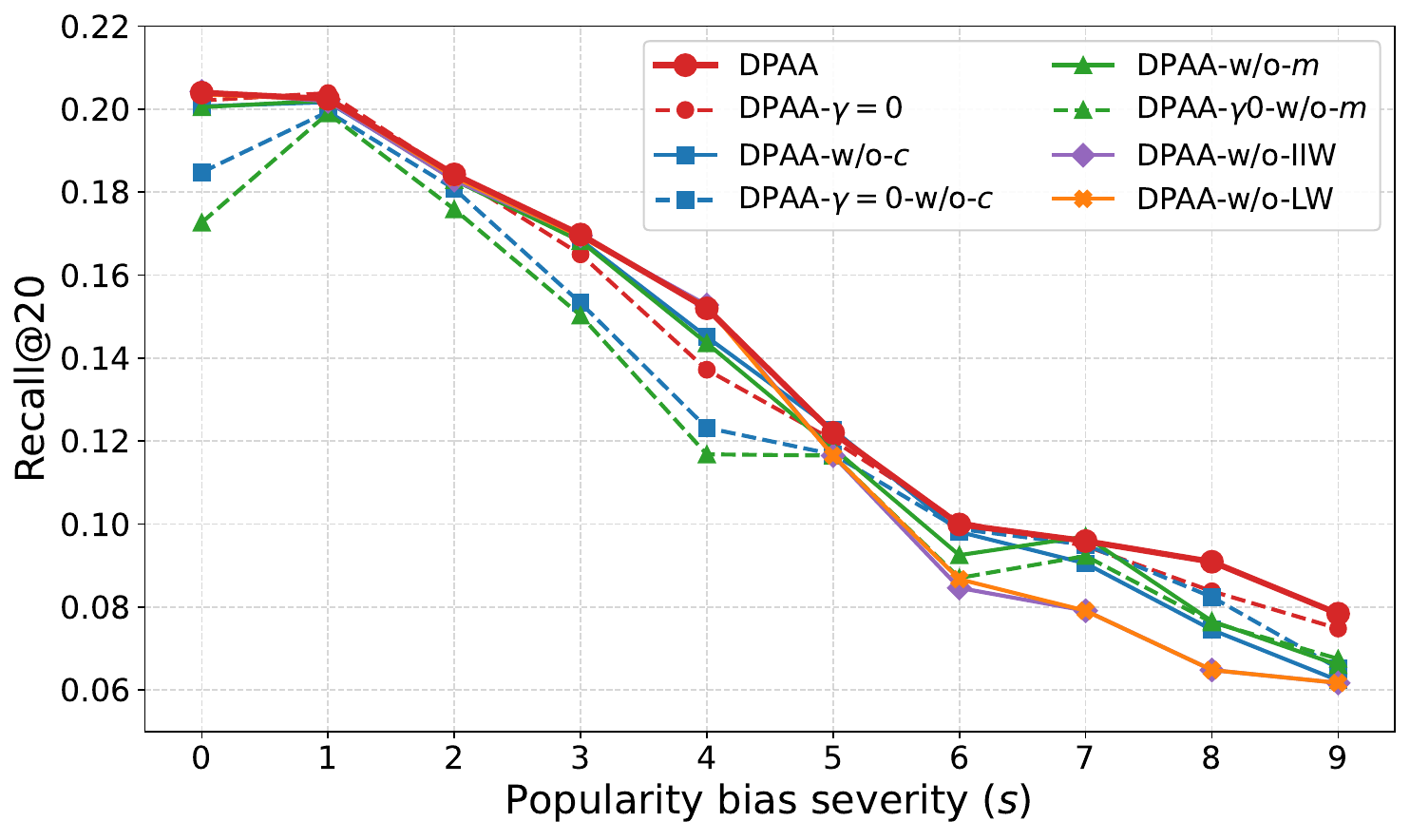}
    \captionsetup{justification=raggedright, margin=0cm}
    \captionsetup{width=0.47\textwidth}
    \caption{Performance comparison of DPAA variants for Recall@20 on the KuaiRec dataset by varying levels of popularity bias severity ($\bm{s}$), showing the contribution of different components.}
    \label{fig:dpaa_variants}
\end{wrapfigure}

We analyze the contribution of different components in DPAA with the KuaiRec semi-synthetic clicks under varying levels of popularity bias severity ($s$). We show the results in Figure~\ref{fig:dpaa_variants}. By default, DPAA uses $\gamma=1$, meaning that IIW is applied only at the first layer.

First, comparing DPAA, which applies IIW only in the initial layer, with DPAA-$\gamma{=}0$, where IIW is applied to all layers, we observe that DPAA performs better in most cases. This suggests that restricting IIW to the initial layer is generally effective, since applying it across all layers may impose overly aggressive correction and remove useful higher-order collaborative signals.

Next, we examine the role of combining pre-trained ($m$) and current ($c$) embeddings for IIW estimation. To do so, we consider two variants: one that excludes the pre-trained model estimate and another that excludes the current model estimate when computing IIW. Removing either the pre-trained estimate (DPAA-w/o-$m$ and DPAA-$\gamma{=}0$-w/o-$m$) or the current estimate (DPAA-w/o-$c$ and DPAA-$\gamma{=}0$-w/o-$c$) generally degrades performance compared with DPAA. This indicates that relying solely on either pre-trained or current embeddings is insufficient, whereas combining them through a smooth transition from the pre-trained estimate to the dynamic model estimate provides a more stable and accurate IIW estimation.

Finally, removing the other key components, IIW and LW, leads to clear performance drops. DPAA-w/o-IIW typically underperforms DPAA, highlighting the importance of interaction-level weighting in reducing popularity bias. Similarly, DPAA-w/o-LW shows degraded performance compared to DPAA, indicating that layer-wise weighting is crucial for mitigating shallow-layer dominance and preserving long-range collaborative signals.

Overall, these results demonstrate that both IIW and LW are essential components of DPAA. They also show that combining pre-trained and current embeddings through a smooth transition is important for reliable IIW estimation and improved performance. Together, these findings highlight the importance of all key components of our method in reducing popularity bias and improving recommendation performance. We also provide the results in Table~\ref{tab:dpaa_variants}.

\begin{table*} [hbtp]
\setlength{\tabcolsep}{2.0pt}
\fontsize{7.4}{8}\selectfont
\centering
\caption{Performance comparison of DPAA variants for Recall@20 on the KuaiRec dataset by varying levels of popularity bias severity ($\bm{s}$), showing the contribution of different components. Bold and underline indicate the best and second-best results.}
\label{tab:dpaa_variants}
\begin{tabular}{ccccccccc}
\toprule
$s$ & DPAA & DPAA-$\gamma{=}0$ & DPAA-w/o-$c$ & DPAA-$\gamma{=}0$-w/o-$c$ & DPAA-w/o-$m$ & DPAA-$\gamma{=}0$-w/o-$m$ & DPAA-w/o-IIW & DPAA-w/o-LW \\
\midrule
0.0  & \underline{0.2040} & 0.2021 & 0.2007 & 0.1848 & 0.2005 & 0.1727 & \textbf{0.2043} & 0.2036 \\
1.0  & \underline{0.2025} & \textbf{0.2039} & 0.2017 & 0.1994 & 0.2021 & 0.1990 & 0.2019 & 0.2024 \\
2.0  & \textbf{0.1843} & \underline{0.1841} & 0.1833 & 0.1808 & 0.1829 & 0.1759 & 0.1829 & 0.1836 \\
3.0  & \textbf{0.1698} & 0.1650 & 0.1685 & 0.1533 & 0.1682 & 0.1503 & \underline{0.1695} & 0.1693 \\
4.0  & 0.1520 & 0.1372 & 0.1451 & 0.1232 & 0.1436 & 0.1168 & \textbf{0.1528} & \underline{0.1524} \\
5.0  & \underline{0.1220} & 0.1202 & \textbf{0.1228} & 0.1168 & 0.1183 & 0.1165 & 0.1165 & 0.1165 \\
6.0  & \underline{0.0999} & \textbf{0.1005} & 0.0981 & 0.0988 & 0.0925 & 0.0870 & 0.0846 & 0.0867 \\
7.0  & \underline{0.0959} & 0.0947 & 0.0905 & 0.0950 & \textbf{0.0970} & 0.0923 & 0.0791 & 0.0791 \\
8.0  & \textbf{0.0909} & \underline{0.0838} & 0.0745 & 0.0824 & 0.0766 & 0.0763 & 0.0648 & 0.0648 \\
9.0  & \textbf{0.0784} & \underline{0.0748} & 0.0623 & 0.0652 & 0.0660 & 0.0675 & 0.0617 & 0.0617 \\
\bottomrule
\end{tabular}
\end{table*}

\subsubsection{Sensitivity analysis of the pre-trained model} \label{sec:pre-train_sensitivity}
We further examine the sensitivity of DPAA to the quality of the pre-trained model used to estimate IIW. We consider LightGCN models trained for 20 epochs, 50 epochs, and until full convergence, where fewer epochs generally correspond to less well-trained embeddings and thus lower-quality pre-trained embeddings. As shown in Table~\ref{tab:pretrain_sensitivity}, DPAA remains reasonably robust even when the quality of the pre-trained model is lower. On Coat, Recall@20 and NDCG@20 remain unchanged at 0.2565 and 0.2065, respectively, across 20 epochs, 50 epochs, and full convergence.  On KuaiRec, Recall@20 improves from 0.0280 at 20 epochs to 0.0301 at 50 epochs and 0.0349 with full convergence; NDCG@20 follows a similar improvement from 0.0507 to 0.0530 and 0.0586. On Yahoo! R3, the differences are smaller, with Recall@20 increasing from 0.1442 to 0.1453 and 0.1479, and NDCG@20 increasing from 0.0653 to 0.0655 and 0.0659. Overall, better-quality pre-trained models generally improve DPAA's performance, while lower-quality pre-trained embeddings do not cause substantial degradation. Therefore, these results demonstrate that DPAA is not highly sensitive to the quality of the pre-trained model. This robustness is partly because IIW does not rely solely on the pre-trained model, but also incorporates the interaction signal learned by the current model during training.

\begin{table*} [htpb]
   \fontsize{9}{8}\selectfont
   \setlength{\tabcolsep}{6.0pt}
   \captionsetup{justification=raggedright, width=\textwidth}
   \caption{Sensitivity analysis of DPAA to the quality of the pre-trained model across different datasets. We use pre-trained LightGCN models after 20 epochs, 50 epochs, and full convergence to estimate the IIW in our DPAA method. Bold and underline indicate the best and second-best results.}
   \centering
    \begin{tabular}{ccccccc}
    \toprule
     \multirow{2}{*}{Pre-trained model}
     & \multicolumn{2}{c}{Coat}
     & \multicolumn{2}{c}{KuaiRec}
     & \multicolumn{2}{c}{Yahoo! R3} \\
     \cmidrule(r){2-3} \cmidrule(r){4-5} \cmidrule(r){6-7}
        & Recall & NDCG
        & Recall & NDCG
        & Recall & NDCG \\
    \midrule
    20 epochs
        & \textbf{0.2565} & \textbf{0.2065}
        & 0.0280 & 0.0507
        & 0.1442 & 0.0653 \\
    50 epochs
        & \textbf{0.2565} & \textbf{0.2065}
        & \underline{0.0301} & \underline{0.0530}
        & \underline{0.1453} & \underline{0.0655} \\
    Fully converged
        & \textbf{0.2565} & \textbf{0.2065}
        & \textbf{0.0349} & \textbf{0.0586}
        & \textbf{0.1479} & \textbf{0.0659} \\
    \bottomrule
    \end{tabular}
    \label{tab:pretrain_sensitivity}
\end{table*}

\subsubsection{Parameter sensitivity analysis}
\label{sec:ablation_parameter}
To examine the sensitivity of DPAA to its key hyperparameters, we conduct additional experiments by varying the initial residual connection parameter $\delta$, the layer-wise weighting parameter $\eta$, and the IIW transition parameter $C$. We evaluate performance on the Coat, KuaiRec, and Yahoo! R3 datasets. The parameter search ranges follow the implementation settings described in Section~\ref{app:impl_details}.

\begin{figure} [htpb]
  \centering
  \captionsetup{justification=raggedright, margin=0cm}
  \captionsetup[subfigure]{labelfont=normalfont, textfont=normalfont}

  \begin{subfigure}{0.32\textwidth}
    \includegraphics[width=\linewidth]{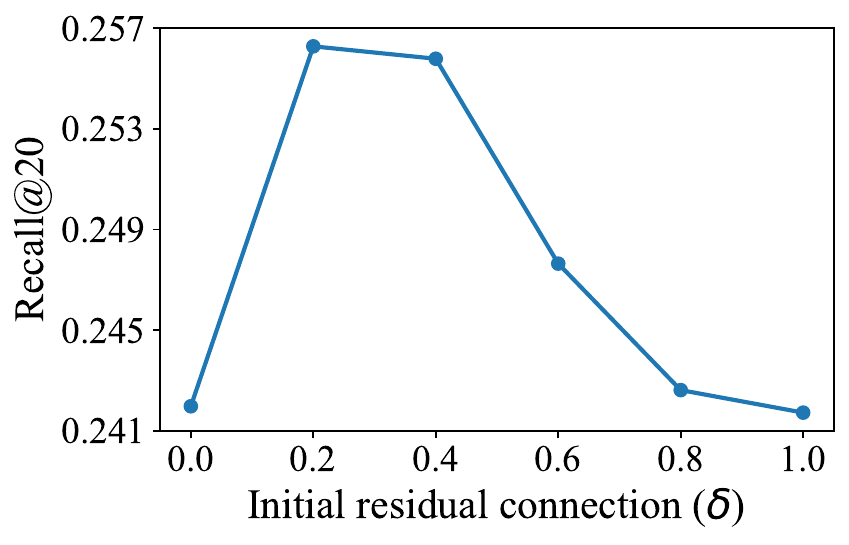}
    \caption{Coat dataset}
  \end{subfigure}
  \begin{subfigure}{0.32\textwidth}
    \includegraphics[width=\linewidth]{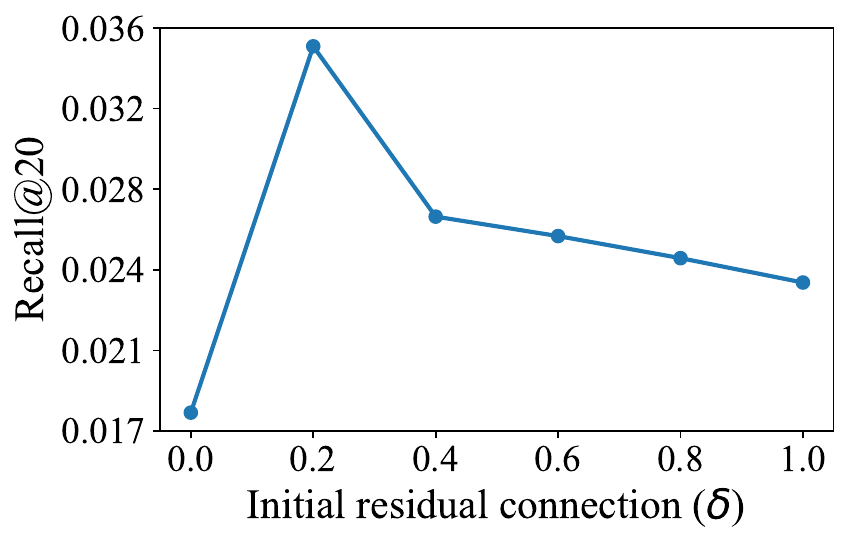}
    \caption{KuaiRec dataset}
  \end{subfigure}
  \begin{subfigure}{0.32\textwidth}
    \includegraphics[width=\linewidth]{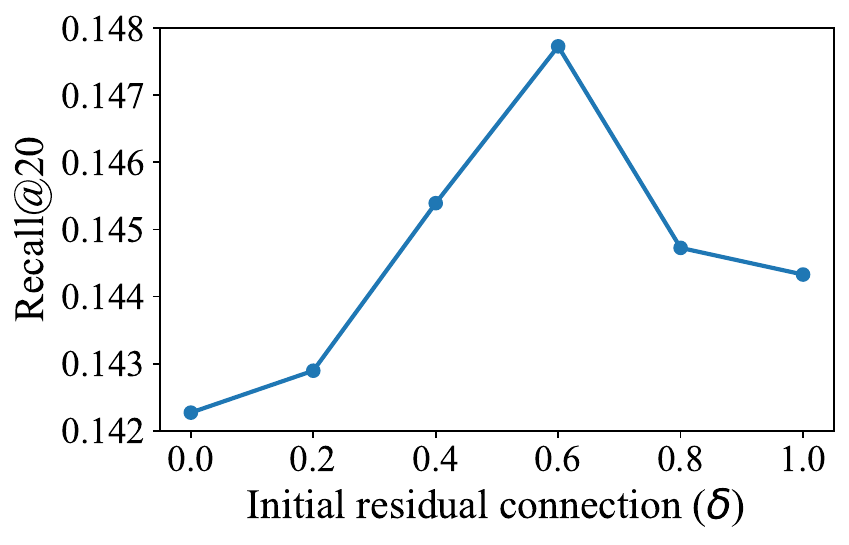}
    \caption{Yahoo! R3 dataset}
  \end{subfigure}

  \captionsetup{width=1.0\textwidth}
  \caption{Effect of varying the \textit{initial residual connection}  ($\bm{\delta}$) across different datasets.\label{fig:residual_coefficient_variations_recall}}
\end{figure}

\textbf{Effect of the initial residual connection.}
Figure~\ref{fig:residual_coefficient_variations_recall} shows the effect of varying $\delta$, which controls the contribution of the initial node embeddings relative to neighborhood aggregation. The results indicate that retaining a moderate amount of initial embedding information is generally beneficial. On Coat, Recall@20 increases substantially from $\delta=0$ and reaches its highest value around $\delta=0.2$, after which performance gradually decreases. KuaiRec exhibits a similar pattern, with the best performance at $\delta=0.2$. On Yahoo! R3, performance improves up to a larger value of $\delta$, reaching its maximum around $\delta=0.6$. These results suggest that the optimal amount of residual information depends on the characteristics of each dataset. In particular, preserving initial embeddings can help prevent over-smoothing while still allowing message passing to incorporate higher-order collaborative signals.

\begin{figure} [htpb]
  \centering
  \captionsetup{justification=raggedright, margin=0cm}
  \captionsetup[subfigure]{labelfont=normalfont, textfont=normalfont}

  \begin{subfigure}{0.32\textwidth}
    \includegraphics[width=\linewidth]{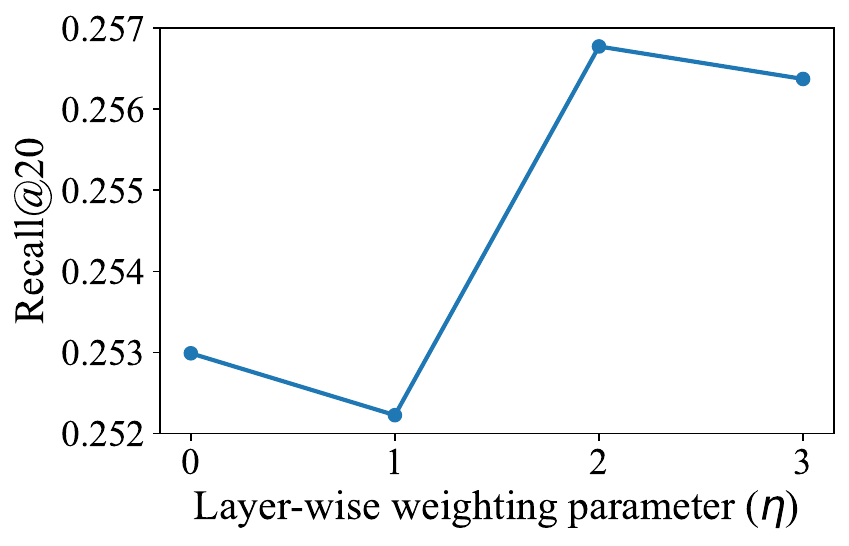}
    \caption{Coat dataset}
  \end{subfigure}
  \begin{subfigure}{0.32\textwidth}
    \includegraphics[width=\linewidth]{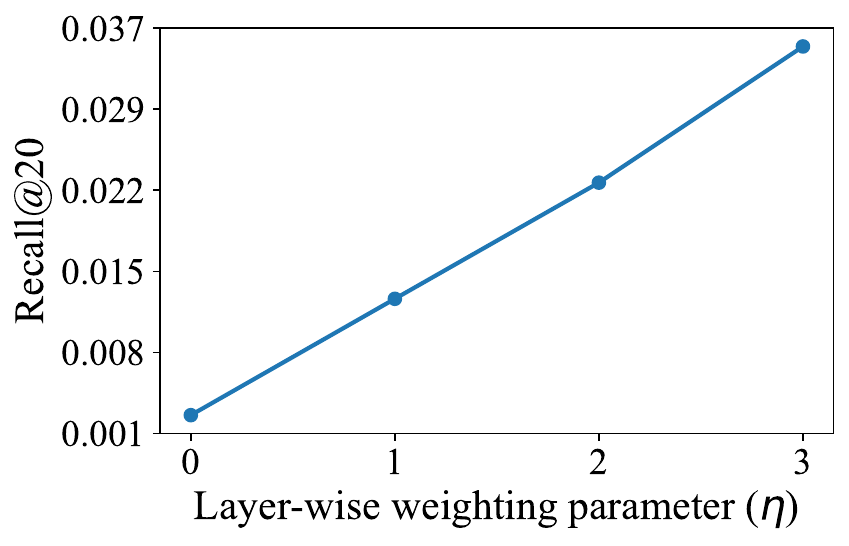}
    \caption{KuaiRec dataset}
  \end{subfigure}
  \begin{subfigure}{0.32\textwidth}
    \includegraphics[width=\linewidth]{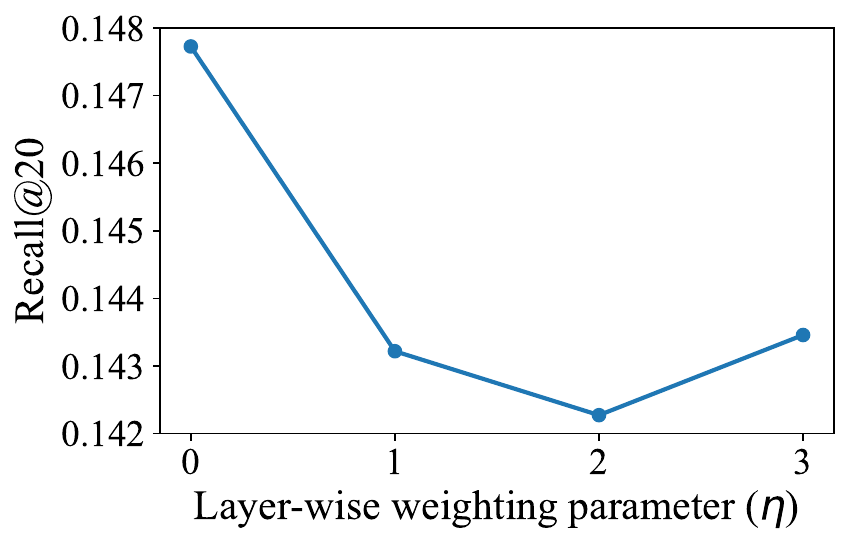}
    \caption{Yahoo! R3 dataset}
  \end{subfigure}

  \captionsetup{width=1.0\textwidth}
  \caption{Effect of varying the layer-wise weighting parameter ($\bm{\eta}$) across different datasets.\label{fig:layer_weight_variations_recall}}
\end{figure}

\textbf{Effect of layer-wise weighting.}
Figure~\ref{fig:layer_weight_variations_recall} evaluates the layer-wise weighting parameter $\eta$, which determines how strongly higher-order representations are emphasized. The effect varies across datasets. On Coat, performance is highest at $\eta=2$, indicating that emphasizing deeper layers can improve recommendation quality even under relatively mild popularity bias. KuaiRec shows a stronger benefit from increasing $\eta$, with Recall@20 increasing consistently across the evaluated range and reaching its highest value at $\eta=3$. This is consistent with the stronger popularity skew in KuaiRec, where higher-order collaborative signals can help reduce the dominance of shallow, popularity-driven interactions. In contrast, Yahoo! R3 achieves its best performance at $\eta=0$, suggesting that uniformly weighting the layers is preferable in this sparse setting.

\begin{figure} [htpb]
  \centering
  \captionsetup{justification=raggedright, margin=0cm}
  \captionsetup[subfigure]{labelfont=normalfont, textfont=normalfont}

  \begin{subfigure}{0.32\textwidth}
    \includegraphics[width=\linewidth]{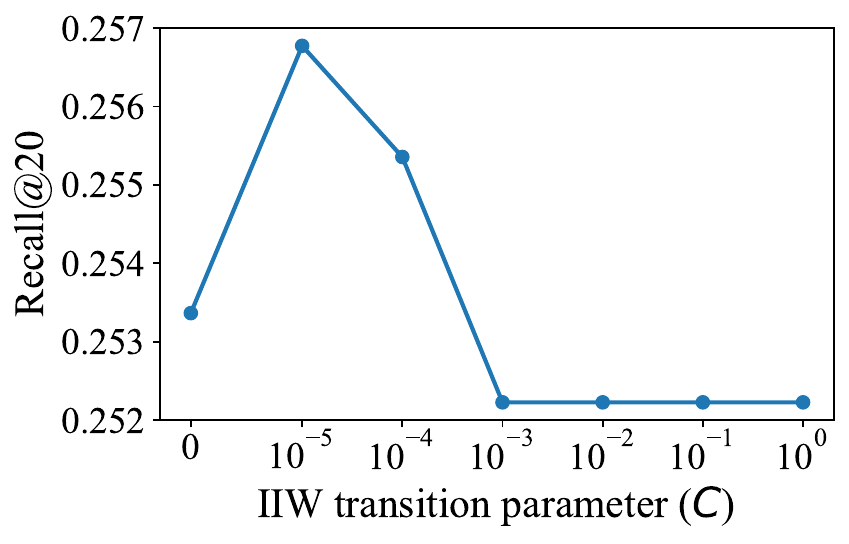}
    \caption{Coat dataset}
  \end{subfigure}
  \begin{subfigure}{0.32\textwidth}
    \includegraphics[width=\linewidth]{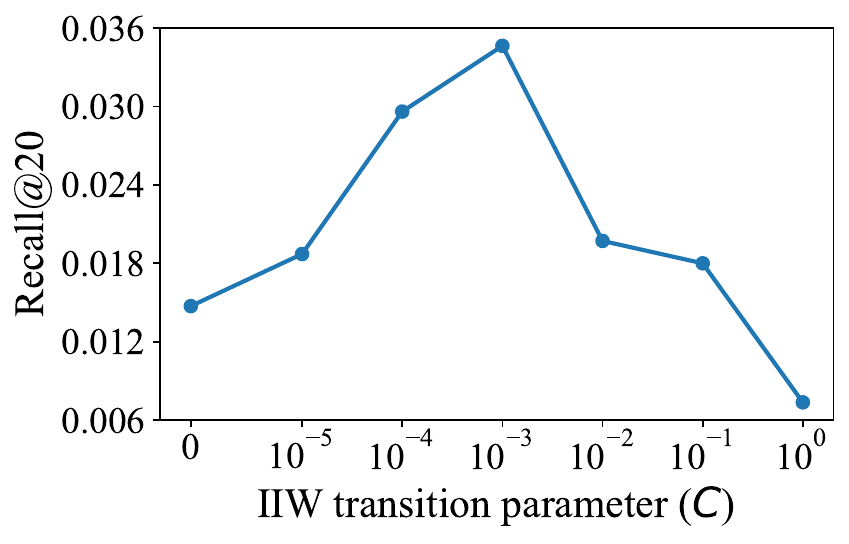}
    \caption{KuaiRec dataset}
  \end{subfigure}
  \begin{subfigure}{0.32\textwidth}
    \includegraphics[width=\linewidth]{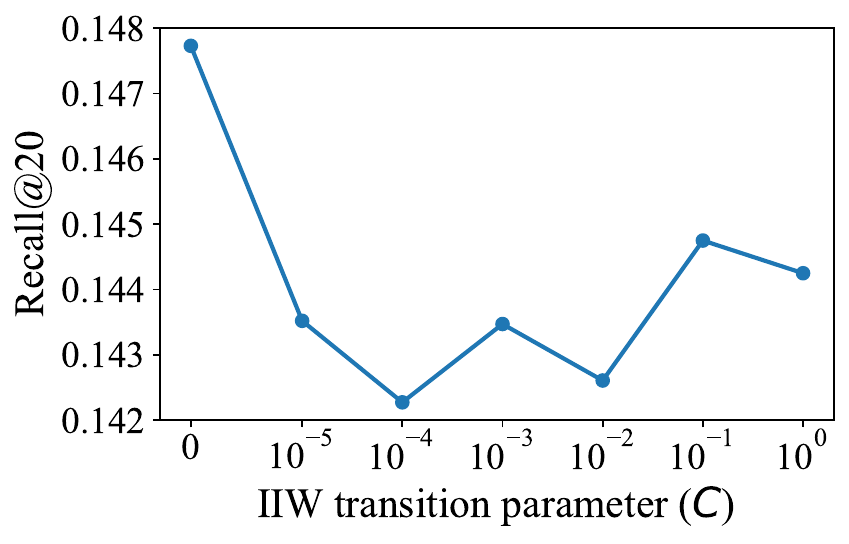}
    \caption{Yahoo! R3 dataset}
  \end{subfigure}

  \captionsetup{width=1.0\textwidth}
  \caption{Effect of varying the IIW transition parameter ($\bm{C}$) across different datasets.\label{fig:C_variations_recall}}
\end{figure}

\textbf{Effect of the IIW transition parameter.}
Figure~\ref{fig:C_variations_recall} studies $C$, which controls the transition between the pre-trained and current-model embeddings used to estimate IIW. Performance varies for different values of $C$ on different datasets. On Coat, a small but nonzero value of $C$ provides the strongest performance, with $C=10^{-5}$ selected as the best configuration. On KuaiRec, performance improves as $C$ increases from very small values and reaches its maximum around $C=10^{-3}$, after which it decreases. This suggests that combining stable pre-trained representations with adaptive current-model representations can provide a more reliable interaction-strength estimate than relying exclusively on either source. On Yahoo! R3, $C=0$ performs best, indicating that the pre-trained embeddings are more useful for IIW estimation in this sparse dataset, where current embeddings may be less reliable.

Overall, the results demonstrate that the optimal values of $\delta$, $\eta$, and $C$ vary across datasets. This variation reflects differences in popularity skew and interaction sparsity. The results further support the design choices of DPAA: retaining initial embeddings helps mitigate over-smoothing, layer-wise weighting controls the contribution of higher-order collaborative signals, and the transition mechanism balances stable pre-trained representations with adaptive current-model representations.

\end{document}